
\magnification=\magstep1
\overfullrule=0pt
\def\w{{\cal W}}
\def\sw{{\cal SW}}
\def\bibitem#1{\parindent=8mm\item{\hbox to 6 mm{$\q{#1}$\hfill}}}
\def\n{{\cal N}}

\def \id{{\rm 1 \kern-2.8pt I }}

\def\lb{\lbrack}
\def\rb{\rbrack}
\def\cf#1#2{\lb\Phi^{#1}_{#2}\rb}
\def\cfs#1#2{\lb{{\scriptstyle{\Phi}}^
{\scriptscriptstyle{#1}}_{\scriptscriptstyle{#2}}}\rb}
\def\q#1{\lb#1\rb}
\def\mn{\medskip\smallskip\noindent}
\def\sn{\smallskip\noindent}
\def\bn{\bigskip\noindent}

\font\extra=cmss10 scaled \magstep0 \font\extras=cmss10 scaled 750

\setbox1 = \hbox{{{\extra R}}}
\setbox2 = \hbox{{{\extra I}}}
\setbox3 = \hbox{{{\extra C}}}

\setbox4=\hbox{{{\extra Z}}}
\setbox5=\hbox{{{\extras Z}}}
\setbox6=\hbox{{{\extras z}}}
\def\Z{{{\extra Z}}\hskip-\wd4\hskip 2.5 true pt{{\extra Z}}}

\def\Zed{\hbox{{\extra\Z}}}

      %
\def\BZT{{\rm Z{\hbox to 3pt{\hss\rm Z}}}}
\def\BZS{{\hbox{\sevenrm Z{\hbox to 2.3pt{\hss\sevenrm Z}}}}}
\def\BZSS{{\hbox{\fiverm Z{\hbox to 1.8pt{\hss\fiverm Z}}}}}

\def\BQT{\,\hbox{\hbox to -2.8pt{\vrule height 6.5pt width .2pt
    \hss}\rm Q}}
\def\BQS{\,\hbox{\hbox to -2.1pt{\vrule height 4.5pt width .2pt\hss}$
   \scriptstyle\rm Q$}}
\def\BQSS{\,\hbox{\hbox to -1.8pt{\vrule height 3pt width
   .2pt\hss}$\scriptscriptstyle \rm Q$}}

\def\BCT{\,\hbox{\hbox to -3pt{\vrule height 6.5pt width
     .2pt\hss}\rm C}}
\def\BCS{\,\hbox{\hbox to -2.2pt{\vrule height 4.5pt width .2pt\hss}$
   \scriptstyle\rm C$}}
\def\BCSS{\,\hbox{\hbox to -2pt{\vrule height 3.3pt width
   .2pt\hss}$\scriptscriptstyle \rm C$}}

\def\BHT{{\rm I{\hbox to 5.3pt{\hss\rm H}}}}
\def\BHS{{\hbox{\sevenrm I{\hbox to 4.2pt{\hss\sevenrm H}}}}}
\def\BHSS{{\hbox{\fiverm I{\hbox to 3.5pt{\hss\fiverm H}}}}}

\def\BPT{{\rm I{\hbox to 5pt{\hss\rm P}}}}
\def\BPS{{\hbox{\sevenrm I{\hbox to 4pt{\hss\sevenrm P}}}}}
\def\BPSS{{\hbox{\fiverm I{\hbox to 3pt{\hss\fiverm P}}}}}

\def\BST{\;\hbox{\hbox to -4.5pt{\vrule height 3pt width .2pt\hss}
   \raise 4pt\hbox to -2pt{\vrule height 3pt width .2pt\hss}\rm S}}
\def\BSS{\;\hbox{\hbox to -4.2pt{\vrule height 2.3pt width .2pt\hss}
   \raise 2.5pt\hbox to -4.8pt{\vrule height 2.3pt width .2pt\hss}
   $\scriptstyle\rm S$}}
\def\BSSS{\;\hbox{\hbox to -4.2pt{\vrule height 1.5pt width .2pt\hss}
   \raise 1.8pt\hbox to -4.8pt{\vrule height 1.5pt width .2pt\hss}
   $\scriptscriptstyle\rm S$}}

\def\BFT{{\rm I{\hbox to 5pt{\hss\rm F}}}}
\def\BFS{{\hbox{\sevenrm I{\hbox to 4pt{\hss\sevenrm F}}}}}
\def\BFSS{{\hbox{\fiverm I{\hbox to 3pt{\hss\fiverm F}}}}}

\def\BRT{{\rm I{\hbox to 5.5pt{\hss\rm R}}}}
\def\BRS{{\hbox{\sevenrm I{\hbox to 4.3pt{\hss\sevenrm R}}}}}
\def\BRSS{{\hbox{\fiverm I{\hbox to 3.35pt{\hss\fiverm R}}}}}

\def\BNT{{\rm I{\hbox to 5.5pt{\hss\rm N}}}}
\def\BNS{{\hbox{\sevenrm I{\hbox to 4.3pt{\hss\sevenrm N}}}}}
\def\BNSS{{\hbox{\fiverm I{\hbox to 3.35pt{\hss\fiverm N}}}}}
\def\BN{{\mathchoice{\BNT}{\BNT}{\BNS}{\BNSS}}}
\def\BAT{\hbox{\raise1.8pt\hbox{\sevenrm/}{\hbox to 4pt{\hss\rm A}}}}
\def\BAS{\hbox{\raise1.4pt\hbox{\fiverm/}
{\hbox to 3pt{\hss\sevenrm A}}}}
\def\BASS{\hbox{\raise1.4pt\hbox{\fiverm/}
{\hbox to 3pt{\hss\sevenrm A}}}}

\def\upin{\hbox{${\scriptstyle\cup}\hbox to
-2.7pt{\hss\vrule height 3.5pt}$}}
\def\downin{\hbox{${\scriptstyle\cap}\hbox to
-2.7pt{\hss\vrule height3.5pt}$}}
\def\rep{representation }
\def\reps{representations }
\def\alg{algebra }
\def\algs{algebras }

\def\eh{{1\over2}}
\def\dh{{3\over2}}
\def\fh{{5\over2}}
\def\sh{{7\over2}}
\def\nh{{9\over2}}
\def\tauh{{\textstyle{\tau\over2}}}
\def\NSS{\widetilde{NS}}
\def\del#1#2{\delta_{#1,#2}}
\def\supvir{super Virasoro algebra }
\def\vir{Virasoro algebra }

\def\scpmoneh{{\textstyle{{p-1\over2}}}}
\def\scph{{\textstyle{{p\over2}}}}
\def\scqh{{\textstyle{{q\over2}}}}
\def\scso{{\scriptstyle{s=1}}}
\def\section#1{\leftline{\bf #1}
\vskip-7pt
\line{\hrulefill}}
\def\bpz{1}
\def\witten{2}
\def\cardy{3}
\def\ver{4}
\def\moresei{5}
\def\zamo{6}
\def\bai{7}
\def\blg{8}
\def\bal{9}
\def\bou{10}
\def\blm{11}
\def\kauwat{12}
\def\wattswb{13}
\def\wattsrr{14}
\def\rva{15}
\def\wirrep{16}
\def\kwf{17}
\def\jos{18}
\def\blmwir{19}
\def\eh1{20}
\def\sevrin{21}
\def\howcom{22}
\def\cas{23}
\def\ehf{24}
\def\bouwschou{25}
\def\laszlo{26}
\def\dhr{27}
\def\reck{28}
\def\www{29}
\def\wowoB{30}
\def\mus{31}
\def\schell{32}
\def\cap{33}
\def\capsup{34}
\def\Witten{35}
\def\Kastor{36}
\def\pim{37}
\def\roc{38}
\def\roes{39}
\def\goddard{40}
\def\matsuo{41}
\def\Ravanini{40}
\def\mussSS{42}
\def\ralfdip{43}
\def\bowknegt{44}
\def\horstphd{45}
\def\andicorr{46}
\def\pimpriv{47}
\font\HUGE=cmbx12 scaled \magstep4
\font\Huge=cmbx10 scaled \magstep4
\font\Large=cmr12 scaled \magstep3

\font\large=cmr12 scaled \magstep1
%
%
\nopagenumbers
\pageno = 0
\centerline{\HUGE Universit\"at Bonn}
\vskip 10pt
\centerline{\Huge Physikalisches Institut}
\vskip 2.4cm
\centerline{\Large Fusion Algebras of Fermionic}
\vskip 6pt
\centerline{\Large Rational Conformal Field Theories}
\vskip 7pt
\centerline{\Large via a Generalized Verlinde Formula}
\vskip 1.4cm
\centerline{{\large Wolfgang\ Eholzer}\ \  and \ \ {\large Ralf\ H\"ubel}}
\vskip 1.2cm
\centerline{\bf Abstract}
\mn
We prove a generalization of the Verlinde
formula to fermionic rational conformal field
theories. The fusion coefficients of the fermionic
theory are equal to sums of fusion coefficients
of its bosonic projection. In particular,
fusion coefficients of the fermionic theory
connecting two conjugate Ramond fields
with the identity are either one or two.
Therefore, one is forced to weaken
the axioms of fusion algebras for fermionic theories.
We show that in the special case
of fermionic $\w(2,\delta)$-algebras
these coefficients are given by the
dimensions of the irreducible representations of
the horizontal subalgebra on the highest weight.
As concrete examples we discuss fusion algebras of rational
models of fermionic $\w(2,\delta)$-algebras including
minimal models of the $N=1$ super Virasoro algebra as well as
$N=1$ super $\w$-algebras $\sw(\dh,\delta)$.
\vskip 20pt
\noindent
\vfill
\settabs \+&  \hskip 110mm & \phantom{XXXXXXXXXXX} & \cr
\+ & Post  address:                      & BONN-HE-93-05   & \cr
\+ & Nu{\ss}allee 12                     & hep-th/9307031  & \cr
\+ & D-53115 Bonn                        & Bonn University & \cr
\+ & Germany                             & July 1993      & \cr
\+ & e-mail:                             & ISSN-0172-8733  & \cr
\+ & eholzer@mpim-bonn.mpg.de            &                 & \cr
\+ & ralf@avzw01.physik.uni-bonn.de      &                 & \cr
\eject
\pageno=1
\footline{\hss\tenrm\folio\hss}
\section{1. Introduction}
\bn
In the last years two-dimensional conformally invariant
quantum field theories
have found wide applications in various fields of
physics and mathematics such as statistical mechanics, string
theory, knot theory, number theory and the classification
of 3-manifolds $\q{\bpz-\moresei}$.
As was shown by A.A.\ Belavin, A.M.\ Polyakov and A.B.\ Zamolodchikov
in 1984 $\q{\bpz}$, rational conformal field theories
(RCFTs) are of particular interest because all $n$-point functions can
be calculated explicitly (at least in principle).
Therefore, the classification of all RCFTs is one of the outstanding
problems in mathematical physics. To this end many different
approaches have been developed $\q{\zamo-\laszlo}$. One direction is
the purely algebraic approach where one considers abstract observable
algebras ($C^*$-algebras) and endomorphisms thereof $\q{\dhr}$.
Here, the fusion rules appear naturally if one decomposes the product
of two endomorphisms into the irreducible ones $\q{\dhr}$ which
can e.g.\ be calculated under some additional assumptions
using algebraic $K$-theory $\q{\reck}$.
A more concrete ansatz follows from the assumption
that all rational models can be described as minimal
models of an extension of the conformal algebra. The investigation
of these $\w$-symmetries in conformal field theory is still one of the
main streams of research in this field of mathematical physics
$\q{\zamo-\bouwschou}$. In this approach one tries
to construct an algebra of local fields
and searches for rational models by investigating representation
theory $\q{\blm-\howcom}$.
The fusion rules describing the interactions of a RCFT
are directly linked to the modular properties of the characters of the
chiral algebra via the Verlinde formula $\q{\ver,\moresei}$.
\sn
Recently, new methods tried to deal directly with the fusion
algebras $\q{\cas}$ (cf.\ references therein) or the fusion algebras
induced by representations of the modular group
$\q{\ehf}$ via the Verlinde formula. In a RCFT a
representation of the modular group is given by the natural
action of $SL_2(\Zed)$ on the characters of the
highest weight representations (HWRs) of the
(maximally extended) chiral symmetry algebra $\w$
underlying the RCFT $\q{\cardy,\www}$.
An important tool in this approach is the famous Verlinde formula
$\q{\ver,\moresei}$ which establishes the connection between the
representation matrix $S$ of the modular transformation
$\tau\rightarrow -{1\over\tau}$ and the fusion coefficients themselves.
In the case of bosonic extended symmetry algebras,
several ans\"atze using the Verlinde formula
led to interesting results $\q{\cas,\wirrep,\ehf,\wowoB}$.
The main goal of this article is to establish a generalized
Verlinde formula which describes the fusion in
all sectors of fermionic theories. This generalization of the
Verlinde formula reproduces the correct sector structure of the
fusion algebra $\q{\mus}$.
We show that the fusion algebras given by the generalized
Verlinde formula can be obtained from the fusion algebras of the
corresponding bosonic projections applying `simple current'
arguments.
Such `simple current' arguments have been first
proposed by A.N.\ Schellekens and S.\ Yankielowicz $\q{\schell}$.
Furthermore, we investigate the representations of the horizontal
subalgebras on the highest weights in the Ramond sector. For
fermionic $\w(2,\delta)$-algebras their dimensions are encoded in
the corresponding fusion algebra.
\mn
This paper is organized as follows. In section $2$ we discuss
some fundamental properties of fermionic RCFTs and present the
explicit form of the generalized Verlinde formula. Here, the
main statements of our analysis are formulated. In the next
section we prove the generalized Verlinde formula
under certain assertions. In section $4$
we apply the formula to the case of $\w$-algebras
with one additional fermionic generator of conformal
dimension $\delta\ge{5\over2}$ ($\w(2,\delta)$).
We proceed with a discussion of the fusion algebras
of the (unitary as well as non-unitary) minimal models
of the $N=1$ super Virasoro algebra in section $5$.
Section $6$ contains the most complicated examples,
namely $N=1$ super $\w$-algebras with
two generators ($\sw(\dh,\delta)$).
Finally, we draw conclusions from our
results and point out some open questions. Two concrete examples
where the fusion algebras of the bosonic projection
of the $N=1$ super Virasoro algebra using the ordinary
Verlinde formula are presented in the appendix.
\bn\sn
\section{2. Generalized Verlinde formula for fermionic RCFTs}
\bn
Let $\cal R$ be a fermionic rational conformally invariant
quantum field theory with (maximally extended) chiral
symmetry algebra $\w$ containing the Virasoro algebra.
One has to distinguish between the Neveu-Schwarz
sector ($NS$) and the Ramond sector ($R$)
because the symmetry algebra contains bosonic and fermionic fields.
We denote the $\w$-primary fields and the corresponding highest weight
representations of the symmetry algebra enlarged by
the (involutive) `chirality'-operator
$\Gamma = (-1)^{\cal F}$ ($\cal F$ is the fermion number operator)
in the two sectors by
$$\vbox{\settabs\+\indent&
$\phi^{NS}_i  \ \ \leftrightarrow \ \ {\cal H}_i^{NS}$\hskip 1cm&\cr
\+&$\phi^{NS}_i  \ \ \leftrightarrow \ \ {\cal H}_i^{NS}$&
${\rm for } \ \ i \in {\cal I}_{NS}$\cr
\+&$\phi^{R }_j \ \ \ \ \leftrightarrow \ \ {\cal H}_j^{R }$&
${\rm for } \ \ j \in {\cal I}_{R}$\cr}
\eqno{(2.1)}$$
where ${\cal H}_1^{NS}$ is the vacuum representation.
The corresponding characters are defined by:
$$\eqalign{
\chi^{NS}_k &:=
{\rm tr}_{{\cal H}^{NS}_k}\bigl( q^{L_0-c/24} \bigr)\cr
\chi^{R}_k &:=
{\rm tr}_{{\cal H}^{R}_k}\bigl( q^{L_0-c/24} \bigr).\cr
}\eqno{(2.2a)}
$$
\mn
The sector structure of such a fermionic theory is
reflected by the modular properties of the characters.
Let $T$\ ($S$) be the \rep matrix of the modular
transformation $\tau\rightarrow\tau+1$
\ ($\tau\rightarrow-{1\over\tau}$) for the representation
of the modular group given by its natural action on the characters.
Because the span of the characters in the $NS$ sector
is invariant only under the subgroup of the modular group
generated by $T^2$ and $S$, it is useful to define a
third (physically irrelevant) sector $\NSS$
(see e.g. $\q{\cap,\capsup}$) in order to obtain a representation
space of the full modular group:
$$\eqalign{
\chi^{\NSS}_k &:=
e^{-2\pi i(h(\phi^{NS}_k)-c/24)}\ \ T\chi^{NS}_k\cr
              &\phantom{:}= {\rm tr}_{{\cal H}^{NS}_k}
            \bigl((-1)^{\cal F} q^{L_0-c/24} \bigr).\cr
}\eqno{(2.2b)}
$$
The modular transformation $TST$ intertwines between
the $NS$ and the $R$ sectors. Together with the $R$ sector,
which is invariant under $T$ and $S T^2 S$, these
three sectors have the structure of a $SL_2(\Zed)$ module.
The `horizontal' subalgebra is defined as the subalgebra
consisting of the zero modes of all fields in $\w$ and
the `chirality'-operator $\Gamma$.
We stress that in our convention the characters $\chi^R$ begin with
$q^{h-{c\over{24}}}(d+\dots)$, where $d$ is the dimension of
the $L_0$ eigenspace ${\cal V}_0$ to the lowest eigenvalue $h$
in the \rep module. A highest weight vector is
an eigenvector of a maximal number of zero-modes of fields
in the bosonic part of the horizontal subalgebra which commute in
${\cal V}_0$. Therefore, $d$ is in general
greater than one because zero-modes of fermionic fields act
nontrivially on the highest weight vector generating ${\cal V}_0$.
In the following we call the characters $(2.2)$ `energy'
characters. In particular,
we do not include a factor $\sqrt{2}$ in
the $R$ characters like in ref.\ $\q{\capsup}$.
It may happen that some of the irreducible \rep modules
of the \alg $\w$ are degenerate and have equal energy
characters. In this case the sector structure of the \rep of
the modular group is respected only if one considers
the energy characters and identifies the degenerate ones.
\vskip 0cm\noindent
In analogy to the $\NSS$ sector one can define
characters in the $\widetilde{R}$ sector by
$$\eqalign{
\chi^{\widetilde{R}}_k &:= {\rm tr}_{{\cal H}^{R}_k}
                   \bigl((-1)^{\cal F} q^{L_0-c/24} \bigr).
}\eqno{(2.2c)}$$
This sector is invariant under the action of the modular group,
and the $\widetilde{R}$ characters $(2.2c)$ are constant. These
constants can be identified with Witten indices of the
corresponding highest weight \reps because they indicate
whether the boson-fermion symmetry of the ground state is broken
or not $\q{\Witten, \Kastor}$.
\mn
The transformation properties of the characters lead
to the following form for the \rep matrices of $S$ and
$T$ (we omit the $\widetilde{R}$ sector because of its
modular invariance) $\q{\cap,\capsup}$:
$$\eqalign{
  S &= \pmatrix{ S^{NS\rightarrow NS}& 0& 0 \cr
                 0 & 0 & S^{R\rightarrow\NSS} \cr
                 0 & S^{\NSS\rightarrow R} & 0 \cr} \cr
  T &= \pmatrix{ 0 & T^{\NSS\rightarrow NS} & 0 \cr
                 T^{NS\rightarrow\NSS} & 0 & 0 \cr
                 0 & 0 & T^{R\rightarrow R} \cr }
}\eqno{(2.3)}$$
with $T^{\NSS\rightarrow NS}_{k,l} = T^{NS\rightarrow\NSS}_{k,l}
       = \delta_{k,l}\ e^{2\pi i ( h(\phi^{NS}_k) - c/24 )}$.
\mn
It is well-known that the fusion rules of fermionic fields
respect the sector structure of the theory in the
following way $\q{\mus}$:
$$\eqalign{
\cf{NS}{i}\cf{NS}{j} &= \sum_{k\in {\cal I}_{NS}}
      \bigl({\n_{NS,NS}^{NS}}\bigr)_{i,j}^k \cf{NS}{k} \cr
\cf{R}{l}\cf{R}{m} &= \sum_{i \in {\cal I}_{NS}}
      \bigl({\n_{R ,R}^{NS}}\bigr)_{l,m}^i  \cf{NS}{i} \cr
\cf{R}{l}\cf{NS}{i} &= \sum_{m \in {\cal I}_{R}}
         \bigl({\n_{R ,NS}^{R}}\bigr)_{l,i}^m  \cf{R}{m}.
}\eqno{(2.4)}$$
This can be interpreted as the conservation of an additional
additive $\Zed_2$ charge (the $NS$ sector is neutral and the
$R$ sector carries charge $1$).
Taking into account the sector structure of the fusion rules for
fermionic theories the generalized Verlinde
formula can be written as:
$$\eqalign{
\bigl({\n_{NS,NS}^{NS}}\bigr)_{i,j}^k &= \sum_{n \in {\cal I}_{NS}}
{S^{NS\rightarrow NS}_{n,i}S^{NS\rightarrow NS}_{n,j}
{(S^{NS\rightarrow NS})^{-1}}_{k,n}
\over S^{NS\rightarrow NS}_{n,1}}\cr
\bigl({\n_{R,R}^{NS}}\bigr)_{l,m}^i &=
d_l d_m \sum_{n \in {\cal I}_{NS}}
{S^{R\rightarrow\NSS}_{n,l}S^{R\rightarrow\NSS}_{n,m}
{(S^{NS\rightarrow NS})^{-1}}_{i,n}
\over S^{NS\rightarrow NS}_{n,1}}\cr
\bigl({\n_{R,NS}^{R}}\bigr)_{l,i}^m &=
\bigl({\n_{NS,R}^{R}}\bigr)_{i,l}^m =
{d_l\over{d_m}}\sum_{n \in {\cal I}_{NS}}
{S^{R\rightarrow\NSS}_{n,l}S^{NS\rightarrow NS}_{n,i}
{(S^{R\rightarrow\NSS})^{-1}}_{m,n}
\over S^{NS\rightarrow NS}_{n,1}}
}\eqno{(2.5)}$$
with $i,j,k \in {\cal I}_{NS}$ and $l,m \in {\cal I}_{R}$.
Note that the formula for $\n_{NS,NS}^{NS}$ is the usual
Verlinde formula $\q{\ver,\moresei}$.
The $\{d_l\mid l\in{\cal I}_{R}\}$ are defined in the following
manner. Consider the bosonic projection $\cal{PW}$ of the
symmetry algebra $\w$ and the corresponding rational model
of $\cal{PW}$. The fusion algebra of this rational model
contains a `simple current' of order two describing its
extended symmetry. The orbits under this `simple current'
correspond to the fields in the fusion algebra of $\w$.
Now $d_l$ is defined as the order of the `simple current'
divided by the length of the orbit corresponding to the $l^{th}$
field in the fusion algebra of $\w$. For the fields in the
$NS$ sector $d$ is equal to one (cf.\ section 3).
We combine these integers to a diagonal matrix
{\hbox{$D^R = diag(\{d_l\mid l\in{\cal I}_{R}\})$}}
and define the $D$-matrix as
$$D = \pmatrix{ \id  &  0   &   0   \cr
                  0  & \id  &   0   \cr
                  0  &  0   & D^{R} \cr}
\eqno{(2.6)}$$
The definition $(2.5)$ directly implies that the fusion constants
lead to a well-defined, commutative and associative algebra.
Furthermore, fusion with the identity field acts trivially
in the fusion algebra. However, it is not apparent that the
fusion coefficients defined in $(2.5)$ are positive integers.
In section 3 we will show that this is an immediate consequence
of the fusion coefficients of the bosonic fusion algebra being
positive integers.
\mn
In general, the \rep modules are degenerate and give rise to
a diagonal `multiplicity' matrix $M$ defined as
$$M = \pmatrix{ M^{NS}  &     0    &   0   \cr
                   0    & M^{\NSS} &   0   \cr
                   0    &     0    & M^{R} \cr}
\eqno{(2.7)}$$
with $M^{NS}=M^{\NSS}$. Here the entries of the three diagonal
submatrices are the multiplicities of the respective
representations in the theory. In all known cases degeneracies
can be removed by considering the eigenvalues of some
additional zero modes of bosonic fields ${\cal B}_0$ which
commute with $L_0$. In general, there are several HWRs with
identical energy characters but with different eigenvalues of
the operators in ${\cal B}_0$.
As described above we must use the energy characters
in order to preserve the sector structure so
that we have to deal with the multiplicity matrix $M$.
\mn
Instead of being unitary, $S$ obeys the equation
$$S^{\dagger} H S = H \ \ \ \ \ H = M D^{-1},\eqno{(2.8)}$$
as will be explained in the next section.
One should note that if $S$ and $M$ are known $D$ is fixed by
this equation. Due to $(2.8)$ the `fusion charge
conjugation matrix' $\n^1_{ij}$ is in general not equal
to the usual charge conjugation matrix $C=S^2$
but satisfies $\n^1_{ij} = (M D)_{i,j}$. As we will discuss below, it
is not possible to avoid the $D$-matrix by means of
an extension of the fusion algebra in contrast to the degeneracies
encoded in $M$ which can be resolved (see e.g.\ ref.\ $\q{\schell}$).
Thus, one is forced to weaken
the axioms of fusion algebras of fermionic RCFTs and has to allow
more general fusion charge conjugation matrices.
\mn
Finally, we add some remarks concerning the chirality operator
$\Gamma$ which we always include into the chiral algebra.
In general it is very unphysical to consider representations
with diagonal fermionic operators. However, if the chirality operator
(anticommuting with all fermionic operators) is included into the
symmetry algebra the fermionic operators act non-diagonal and do not
preserve the $\Gamma$-eigenspaces.
Irreducible representations of the fermionic
algebra (including $\Gamma$)
which correspond to orbits of length two in the fusion algebra of
the bosonic projection are also irreducible with respect to the
chiral algebra without $\Gamma$. In contrast, irreducible
representations corresponding to fixed points in the
bosonic fusion algebra are not irreducible with respect to the
chiral algebra without $\Gamma$ but decompose into a direct sum
of two irreducible representations. This may explain the fact that the
fusion coefficients connecting two such conjugate fields with the
identity are equal to two. As we will see in section 4 it is not
possible to resolve these nontrivial fusion coefficients by extending
the fusion algebra.
\mn
In section 4 we verify for fermionic $\w(2,\delta)$-algebras
that the diagonal entries of $D^R$ give exactly the dimensions
of the $L_0$-eigenspaces ${\cal V}_0$ in the respective
representation modules. However, this observation is in
general not valid for symmetry algebras with more than one
fermionic generator like $\sw(\dh,\delta)$-algebras (cf.\ section 6).
\bn\sn
\section{3. Proof of the Generalized Verlinde Formula}
\bn
Assume that we have a bosonic RCFT with characters $\chi_i^{bos}$
($i\in{\cal I}_{bos}$) and unitary $S$-matrix $S^{bos}$.
Furthermore, we suppose for simplicity that ${(S^{bos})}^2 = \id$,
i.e.\ trivial charge conjugation and that there are no degenerate
HWRs so that $N_{i,i}^1 = 1$ ($\forall i\in{\cal I}_{bos}$) is valid.
The bosonic RCFT shall also admit a fermionic extension
of its underlying chiral symmetry algebra. We require
therefore that the fusion algebra
${\cal A}$ obtained from $S^{bos}$ via the (ordinary) Verlinde
formula possesses a `simple current' $\q{J}$ of order
two ($\q{J}\q{J}=\id$) with conformal dimension
$h(\q{J})=\delta\in{\BN+{1\over2}}$ $\q{\schell}$.
Assume furthermore that the fermionic symmetry algebra is obtained
from its bosonic projection by extension with this `simple current'
(this is at least valid for $N=1$ supersymmetric theories and
fermionic $\w(2,\delta)$-algebras).
\bn
The basis elements $\q{i}$ ($i\in{\cal I}_{bos}$) of ${\cal A}$
are organized into orbits of the `simple current' $\q{J}$
of length one (fixed points) or two. These orbits thus
define the multiplets of the extended fermionic RCFT.
Accordingly, the characters of the fermionic theory are defined as
$\chi_i^{fer} := \chi_i^{bos} + \chi_{Ji}^{bos}$.
As shown in ref.\ $\q{\schell}$ it is possible to define
a conserved charge $Q$ on the fusion algebra ${\cal A}$ in the
following way: $Q(\q{i}) := \bigl(h(\q{i}) + h(\q{J}) -
h(\q{J}\q{i})\bigr)\ mod\ 1.$ It obeys the addition rule
$Q(\q{i}\q{j}) = Q(\q{i}) + Q(\q{j}) \ mod \ 1$.
The fields have either charge $0$ or charge ${1\over2}$ with
respect to the `simple current' $\q{J}$. Due to charge
conservation the fusion algebra ${\cal A}$ is $\Zed_2$ graded,
i.e. we have ${\cal A}={\cal A}_0\oplus{\cal A}_{1\over2}$.
Here ${\cal A}_0$ (${\cal A}_{1\over2}$) denotes the subspace
spanned by fields of charge $0\,({1\over2})$.
The fusion rules respect this structure in the following way:
$$\vbox{\settabs 3 \columns
\+${\cal A}_0 \cdot {\cal A}_0 \subset {\cal A}_0$ &
${\cal A}_0 \cdot {\cal A}_{1\over2}\subset {\cal A}_{1\over2}$&
${\cal A}_{1\over2}\cdot {\cal A}_{1\over2}\subset {\cal A}_0$ \cr}
\eqno{(3.1)}$$
We observe that the fields in ${\cal A}_0$ form a subalgebra of
${\cal A}$. It is thus natural to identify the orbits in
${\cal A}_0$(${\cal A}_{1\over2}$) with the fields of the
Neveu-Schwarz (Ramond) sector of the fermionic RCFT
$\q{\schell}$. Note that in ${\cal A}_0$ all orbits have
length two, i.e.\ there are no fixed points in the $NS$ sector.
However, in ${\cal A}_{1\over2}$ fixed points as well as
orbits of length two are possible.
\mn
It is now straightforward to calculate the fusion algebra
for the fermionic RCFT. We just have to perform
the change of basis prescribed by the orbits of $\q{J}$.
The new basis is defined as
$$\q{\hat{\imath}} := {\textstyle{1\over2}}\bigl(\q{i}+
\q{J}\q{i}\bigr)\quad\quad
\alpha(i)=\hat{\imath}\in{\cal I}_{NS}\cup{\cal I}_R,\quad
i\in{\cal I}^{'}_{NS}\cup{\cal I}^{'}_{R}={\cal I}_{bos}$$
where ${\cal I}_{NS}$ and ${\cal I}_R$ label the orbits in
${\cal A}_0$ and ${\cal A}_{1\over2}$ respectively.
Furthermore, ${\cal I}^{'}_{NS}$ and ${\cal I}^{'}_{R}$ label
{\it all} basis elements of ${\cal A}_0$ and ${\cal A}_{1\over2}$.
The action of $\q{J}$ induces a natural surjective map of
${\cal I}^{'}_{NS}({\cal I}^{'}_{R})$ onto
${\cal I}_{NS}({\cal I}_R)$ which we denote by
$\alpha$. The fusion coefficients are then given by:
$$\eqalign{
\n_{\alpha(i),\alpha(j)}^{\alpha(k)}
&= N_{i,j}^k + N_{i,j}^{Jk}
\quad k\in{\cal I}^{'}_{NS},\ i,j\in{\cal I}^{'}_{NS}
\ {\rm or}\ i,j\in{\cal I}^{'}_{R}.\cr
}\eqno{(3.2)}$$
$N_{i,j}^k$ are the fusion coefficients of the bosonic fusion
algebra ${\cal A}$ obtained from $S^{bos}$ and
$\n_{\alpha(i),\alpha(j)}^{\alpha(k)}$
denote the fusion coefficients of the corresponding fusion
algebra of the fermionic RCFT. Note that for
$i\in{\cal I}^{'}_{NS}$ either
$N_{i,\overline{i}}^1$ or $N_{i,\overline{i}}^J$ has to be zero
($\overline{i}$ denotes the field conjugate to $i$) because there
are no fixed points in ${\cal A}_0$. Hence, the fusion charge
conjugation matrix is equal to the usual charge conjugation matrix
in the $NS$ sector.
\mn
In order to prove the generalized Verlinde formula
we have to show that
these fusion coefficients are equal to the coefficients
calculated with formulae $(2.5)-(2.8)$ using the
$S$-matrix $S^{fer}$. We obtain $S^{fer}$ from $S^{bos}$ by
performing the appropriate change of basis on the characters
$\chi_{\alpha(i)}^{fer} := \chi_i^{bos} + \chi_{Ji}^{bos}$.
This implies that $S^{fer}$ in not unitary but obeys
eq.\ $(2.8)$. To formulate it differently we have to show
that the following diagram commutes:
$$
\def\mapright#1{\smash{\mathop{\longrightarrow}\limits^{#1}}}
\def\mapdown#1{\Big\downarrow\rlap{$\vcenter{\hbox{$\scriptstyle#1$}}$}}
\matrix{\noalign{\vskip6pt}S^{bos}&&
\mapright{\chi_{\alpha(i)}^{fer}:=\chi_i^{bos}+\chi_{Ji}^{bos}}&&S^{fer}\cr
\mapdown{\rm Verlinde}&&&&\mapdown{\rm generalized\ Verlinde}\cr
{\rm bosonic\ fusion\ algebra}&&
\mapright{\q{\hat{\imath}}:={\textstyle{1\over2}}
\bigl(\q{i}+\q{J}\q{i}\bigr)}&&{\rm fermionic\ fusion\ algebra}
\cr\noalign{\vskip6pt}}\eqno{(3.3)}$$
Equation $(3.2)$ corresponds to the bottom line of the diagram $(3.3)$.
\sn
Let us first check the result for $\n_{i,j}^k$ with $i,j,k\in{\cal I}_{NS}$.
The $S$-matrix $S^{fer}_{NS\rightarrow NS}$ is given by
$(S^{fer}_{NS\rightarrow NS})_{\alpha(i),\alpha(j)}
= S^{bos}_{i,j} + S^{bos}_{Ji,j}.$
Inserting this in the first relation of $(2.5)$ we obtain (using the fact
that $S^{bos}$ and $S^{fer}_{NS\rightarrow NS}$ are symmetric and orthogonal):
$$\eqalign{
\n_{\alpha(i),\alpha(j)}^{\alpha(k)}&= {1\over2}\sum_{n\in{\cal I}_{NS}^{'}}
{{(S^{bos}_{n,i}+S^{bos}_{n,Ji})(S^{bos}_{n,j}+S^{bos}_{n,Jj})
(S^{bos}_{k,n}+S^{bos}_{k,Jn})}\over{S^{bos}_{n,0}+S^{bos}_{Jn,0}}}\cr
&={1\over4}\sum_{n\in{\cal I}^{'}_{NS}}
{{(S^{bos}_{n,i}+S^{bos}_{n,Ji})(S^{bos}_{n,j}+S^{bos}_{n,Jj})
(S^{bos}_{k,n}+S^{bos}_{k,Jn})}\over{S^{bos}_{n,0}}}\cr
&={1\over4}\bigl(N_{i,j}^k+N_{i,j}^{Jk}+N_{i,Jj}^k+N_{i,Jj}^{Jk}+
N_{Ji,j}^k+N_{Ji,j}^{Jk}+N_{Ji,Jj}^k+N_{Ji,Jj}^{Jk}\bigr)\cr
&= N_{i,j}^k+N_{Ji,j}^k.\cr
}$$
Note that we used the identities $S^{bos}_{Ji,j}=S^{bos}_{i,Jj}$ and
$S^{bos}_{n,0}=S^{bos}_{Jn,0}$ (formula $(4.2)$
in ref.\ $\q{\schell}$). This follows directly from the Verlinde
formula and the defining property of the `simple current' $\q{J}$
(fusion of $\q{J}$ with a basis element of ${\cal A}$ yields
only one (different) basis element).
\sn
Consider now the by far more interesting case of
$\n_{\alpha(i),\alpha(j)}^{\alpha(k)}$ with $i,j\in{\cal I}^{'}_{R}$
(in the case $i\in{\cal I}^{'}_{R},j\in{\cal I}^{'}_{NS}$ one
proceeds in the same way).
For simplicity we treat only the case where $i,j$ correspond to fixed
points in ${\cal A}_{1\over2}$. For fixed points the characters of
the fermionic theory are twice the corresponding bosonic characters.
Due to the chirality operator $\Gamma$ in the underlying chiral
symmetry algebra these characters are indeed the characters of
irreducible HWRs of the fermionic theory.
In particular, for fixed
points the dimension $d$ of the irreducible representation of
the horizontal subalgebra of the symmetry algebra in ${\cal V}_0$ is
two. For HWRs corresponding to orbit length two in
${\cal A}_{1\over2}$ the dimension $d$ equals
one or two depending whether
the conformal dimensions of the fields in the orbit are
different or not. With
$(S^{fer}_{R\rightarrow \widetilde{NS}})_{\alpha(i),\alpha(j)}
= S^{bos}_{i,j}$ \hbox{($i\in{\cal I}_{NS}^{'},
\ j\in{\cal I}_{R}^{'}$}, $j$ corresponding to a fixed point)
we obtain with the second relation of $(2.5)$ the following result
(inserting $d_i=d_j=2$):
$$\eqalign{
\n_{\alpha(i),\alpha(j)}^{\alpha(k)} &= {1\over2}d_i d_j
\sum_{n\in{\cal I}_{NS}^{'}}
{{S^{bos}_{n,i}S^{bos}_{n,j}
(S^{bos}_{k,n}+S^{bos}_{k,Jn})}\over{S^{bos}_{n,0}+S^{bos}_{Jn,0}}}
= \sum_{n\in{\cal I}^{'}_{NS}}{{S^{bos}_{n,i}S^{bos}_{n,j}
(S^{bos}_{k,n}+S^{bos}_{k,Jn})}\over{S^{bos}_{n,0}}}\cr
&= N_{i,j}^k+N_{i,j}^{Jk} = 2N_{i,j}^k.\cr
}$$
The case with orbits of length two in ${\cal A}_{1\over2}$ can be
treated similarly. We see that the diagram $(3.3)$ is indeed
commutative thus prooving our assertion.
\mn
In the remaining part of this paper we apply the generalized
Verlinde formula to various fermionic RCFTs. We start with rational
models of fermionic $\w(2,\delta)$
\algs ($\fh\le\delta\in\BN+{1\over2})$ where it is
believed that the classification is complete.
Then we proceed with minimal models (unitary as well as
non-unitary) of the $N=1$ \supvir ($=\w(2,\dh)$).
In two cases we study the fusion rules of the bosonic
projection of the $N=1$ super Virasoro algebra, described by the
usual Verlinde formula, and demonstrate concretely that the
diagram $(3.3)$ commutes (see appendix).
We continue with rational models of $N=1$ super-$\w$-\algs
$\sw(\dh,\delta)$ ($\delta\ge2$) which are the most
complicated examples because in general both the
multiplicity matrix $M$ and the $D$-matrix are nontrivial.
In particular, we discuss $\sw(\dh,\delta)$-algebras with
vanishing self-coupling constant, where the classification
also seems to be complete. Using the $ADE$-classification and earlier
results about the degeneracies $\q{\eh1}$ of the \rep modules, one
obtains the $D$-matrix for all $\sw(\dh,\delta)$-algebras related to
the $ADE$-classification.
\bn\sn
\section{4. Fermionic $\w(2,\delta)$ \algs}
\mn
In this section we discuss $\w(2,\delta)$-algebras
with one additional generator $W$ with
half-integer conformal dimension $\delta\ge\fh$ $\q{\blm,\rva,\wirrep}$.
Besides the parabolic cases $\q{\pim}$ these algebras exist for Virasoro
minimal values of the central charge which can be understood
by the $ADE$-classification of modular invariant partition functions
$\q{\cap}$. These values of the central charge can be
organized in two series according to the type of the partition function
which is diagonalized by the $\w$-characters. From table $1$
below one can read off these $\w$-characters in terms of
Virasoro characters. Actually, the $\w$-characters are the quantities
which appear in $Z=Z^{NS}+Z^{\NSS}+Z^R$
with their absolute value squared.
The irreducible \reps of the horizontal subalgebra of
$\w$ in the $L_0$-eigenspace ${\cal V}_0$ are
either one or two-dimensional in the $R$ sector.
In the one-dimensional (trivial) \rep $W_0$ is
represented by zero whereas
in the two-dimensional \rep $W_0$ acts non-trivially.
The representation of the horizontal subalgebra in ${\cal V}_0$
is equivalent to the irreducible two-dimensional representation of the
Clifford algebra $Cl(2,0)$ of a two-dimensional euclidian vectorspace:
$$\vbox{\settabs\+\quad&xxxxxxxxxxxxxxxxxxxx&xxxxxxxxxxxxxxxxxxx&\cr
\+& $\{W_0,W_0\} = 2w^2,$ & $\{W_0,\Gamma\} = 0,$
& $\{\Gamma,\Gamma\} = 2.$\cr}$$
The trivial \rep only occurs in those highest weight modules for
which the $\w$-character is a sum of Virasoro characters
with an equal number of terms as in the $NS$ sector. From the \rep
theory of these algebras we see that this is exactly the case for the
HWRs with $w=0$ $\q{\wirrep}$. In the case of $\w(2,\delta)$ \algs
$M$ is equal to the identity matrix $\id$ because
there are no additional `quantum numbers' present (and necessary).
So one can read off the $D$-matrix from table $1$ and verify that
the diagonal entries coincide with the dimensions of the
corresponding vector spaces ${\cal V}_0$.
\bn
\centerline{\vbox{
\hbox{\vbox{\offinterlineskip
\def\tablespace{ height2pt&\omit&&\omit&&\omit&&\omit&\cr }

\def\tablerule{ \tablespace\noalign{\hrule}\tablespace}
\hrule\halign{&\vrule#&\strut\hskip0.1cm\hfil#\hfil\hskip0.1cm\cr
\tablespace
& \alg  && $c(p,q)$  && series && $Z^{NS},Z^R$  &\cr
\tablerule\tablerule
& $\w(2,{{p-4}\over2})$  && $c(p,12)$ && $(E_6,A_{p-1})$ &&
$Z^{NS} =
\sum\limits^{\scpmoneh}_{\scso}
{\scriptstyle\mid\chi_{1,s}+\chi_{7,s}+
                               \chi_{5,s}+\chi_{11,s}\mid}^2 $ &\cr
\tablespace
& \omit  && \omit && \omit &&
$Z^R = \sum\limits^{\scpmoneh}_{\scso}
      {\scriptstyle {1\over2}\mid 2(\chi_{4,s}+\chi_{8,s})\mid}^2  $ &\cr
\tablerule
& $\w(2,{{(p-2)(2k-1)}\over2})$ && $c(p,4k)$ && $(D_{2k+1},A_{p-1})$ &&
$Z^{NS} =\sum\limits^{\scpmoneh}_{\scso}
\sum\limits^{\scriptstyle{2k-1}}_{\scriptstyle{r=1}\atop\scriptstyle{odd}}
{\scriptstyle\mid\chi_{r,s}+\chi_{r,p-s}\mid}^2 $ &\cr
\tablespace
& \omit  && \omit && \omit &&
$Z^R =\!\sum\limits^{\scpmoneh}_{\scso}
                \!{\scriptstyle {1\over2}\mid 2\chi_{2k,s}\mid}^2 +\!
\sum\limits^{\scpmoneh}_{\scso}
\sum\limits^{\scriptstyle{2k-2}}_{\scriptstyle{r=2}\atop\scriptstyle{even}}
\!{\scriptstyle\mid\chi_{r,s}+\chi_{r,p-s}\mid}^2 $ &\cr
\tablespace}\hrule}}
\hbox{\hskip 0.5cm Table 1: partition functions and series of
fermionic $\w(2,\delta)$-algebras $\q{\cap,\blm,\wirrep}$}
}}
\bn\mn
Let us also discuss the parabolic fermionic $\w(2,\delta)$-\algs.
The series is given by the \algs $\w(2,3k)$
existing for $c=1-24k$ with $k\in\BN+{1\over2}$. The following
HWRs are permitted $\q{\wirrep}$
($h_{r,r}=k(r^2-1),\ h_{r,-r}=h_{r,r}+r^2$)
\mn
\settabs\+\indent&$NS$\hskip 1cm
&$h_{{2m+1\over{4k+4}},-{2m+1\over{4k+4}}}$\hskip 1cm&\cr
\+&$NS:$&$h_{{m\over{2k}},{m\over{2k}}}$
&$m = 0,\dots,\lfloor k \rfloor, 2k$\cr
\+& &$h_{{m\over{2k+2}},-{m\over{2k+2}}}$
&$m=1,\dots,\lfloor k+{1\over2}\rfloor$\cr
\+&$R:$&$h_{{2m+1\over{4k}},{2m+1\over{4k}}}$
&$m = 0,\dots,\lfloor k \rfloor-1,\lfloor k \rfloor$\cr
\+& &$h_{{2m+1\over{4k+4}},-{2m+1\over{4k+4}}}$
&$m=0,\dots,\lfloor k+{1\over2}\rfloor-1,
\lfloor k+{1\over2}\rfloor$\cr\noindent
\bn
The modular invariant partition function $Z=Z^{NS}+Z^{\NSS}+Z^R$
is given by the expressions $\q{\pim}$ (the $\theta$ function is
defined below ($4.2$)):
$$\eqalign{
Z^{NS} &= {1\over{\eta^2}}\Bigl(
\mid{\textstyle{1\over2}}(\theta_{0,k}-\theta_{0,k+1})\mid^2 +
\mid{\textstyle{1\over2}}(\theta_{0,k}+\theta_{0,k+1})\mid^2 +
\sum_{m=1}^{\lfloor k \rfloor}\!\mid\theta_{m,k}\mid^2 +
\sum_{m=1}^{\lfloor k+{1\over2}\rfloor}\!\mid\theta_{m,k+1}\mid^2\Bigr)\cr
Z^{R} &= {1\over{\eta^2}}\Bigl(
{\textstyle{1\over2}}\mid\theta_{k,k}\mid^2 +
\,{\textstyle{1\over2}}\mid\theta_{k+1,k+1}\mid^2 +
\sum_{m=0}^{\lfloor k \rfloor-1}\!\mid\theta_{m+{1\over2},k}\mid^2 +
\sum_{m=0}^{\lfloor k+{1\over2}\rfloor-1}\!\mid\theta_{m+{1\over2},k+1}\mid^2
\Bigr)\cr}$$
The part $Z^{\NSS}$ of the partition function is obtained from $Z^{NS}$ by
applying the modular transformation $\tau\rightarrow\tau+1$.
The $\w$-characters are the quantities
which appear in $Z$ with their absolute value squared
divided by the $\eta$-function,
e.g.\ $\chi^{\w}_{vac} =\eta^{-1}{\textstyle{1\over2}}
(\theta_{0,k}-\theta_{0,k+1}).$ From the partition function one can
immediately read off the matrix
$D^R = diag(\{2,2,1,\dots,1\})$ taking into account $M=\id$.
Also in this case the diagonal elements of $D^R$ are equal
to the dimensions of the corresponding spaces ${\cal V}_0$.
The fusion algebra can easily be obtained by the
well-known transformation rules of the theta functions $\q{\pim}$.
\bn
In the rest of this section we discuss the fusion rules emerging from
the generalized Verlinde formula for two series of $\w(2,\delta)$-algebras,
namely those corresponding to $(D_3,A_{p-1})$ and $(E_6,A_{p-1})$.
As an example for the general $(D,A)$-series
we discuss the case $(D_5,A_2)$ and verify that it is not possible to
extend the fusion algebra such that the
fusion charge conjugation becomes equal to the usual charge conjugation.
\mn
Take as the first concrete example
$\w(2,{p-2\over2})$ at the Virasoro minimal value of
the central charge $c=c(p,4) = 1-{3\over2}{(p-4)^2\over p}$ for
odd $p \ge 3$. These algebras are related to the
partition functions of the type $(D_3,A_{p-1})$
(specializing the second case
in the table to $k=1$). In order to give the
characters of these algebras explicitly,
recall the general form of the
Virasoro minimal characters $\q{\roc}$
$$\chi^{Vir}_{r,s}(\tau) = {\eta(\tau)}^{-1}
    \bigl(\theta_{pr-qs,pq}(\tau)-\theta_{pr+qs,pq}(\tau)\bigr)
\ \ \ {\rm for} \ \ 1 \le r \le q-1,\ \ 1 \le s \le p-1,
\eqno{(4.1)}$$
where we have introduced Dedekind's eta function and the Riemann-Jacobi
theta functions
$$ \eta(\tau) = q^{1\over24} \prod_{n \in \BN} (1-q^n) , \ \ \ \ \ \ \
    \theta_{\lambda,k}(\tau) = \sum_{n \in \Zed} q^{(2kn+\lambda)^2\over 4k}
    \ \ \ \ \ \ {\rm with} \ \ \ q = e^{2\pi i \tau}. \eqno{(4.2)}$$
The $\w$-algebra characters can be expressed in terms of
the Virasoro minimal characters as $\q{\rva}$
$$\eqalign{
NS: \ \ \ \chi^{\w,NS}_i &= \chi_{1,i} + \chi_{1,p-i} \ \ \ \
     {\rm for} \ \ i \in {\cal I}_{NS} =
               \{1,...,{\scriptstyle{p-1\over2}} \} \cr
R: \ \ \ \ \chi^{\w,R}_i \ &= 2\chi_{2,i}
   \ \ \ \ \ \ \ \ \ \ \ \ \ \
     {\rm for} \ \ i \in {\cal I}_{R} =
                \{1,...,{\scriptstyle{p-1\over2}} \}.
}\eqno{(4.3)}$$
Using the arguments given above, we infer
from $(4.3)$ that the matrix $D^R$ is equal to $2\id$.
Rewriting the modular transformation matrix $S$ of the Virasoro minimal
models $\q{\roc}$ in the $\w$-algebra character basis, one obtains
(for $S^{NS,NS}$ see $\q{\rva}$)
$$\eqalign{
S^{NS,NS}_{i,j} &= {2\over \sqrt{p}}
                   (-1)^{i+j+1+{p-1\over2}+{\lfloor{p+1\over4}\rfloor}}
            {\rm sin}\bigl({4\pi i j \over p}\bigr)
\ \ \ \ \ \ \ \ \ \
                   {\rm{for}} \ \ i,j \in {\cal I}_{NS} \cr
S^{\NSS,R}_{i,l} \ \ &= {1\over2}S^{R,\NSS}_{l,i}
                     =  {2\over{\sqrt{2p}}}(-1)^{l+1+{p-1\over2}}
            {\rm sin}\bigl({4\pi i l \over p}\bigr) \ \ \ \ \ \ \!
{\rm{for}} \ \ i \in {\cal I}_{NS},\ l \in {\cal I}_{R}.
}\eqno{(4.4)}$$
Inserting this result into $(2.5)$ gives the fusion algebra.
This fusion algebra is isomorphic to the fusion algebra of the
Virasoro minimal model with $c=c(2,p)$ tensor an element $\omega$
with $\omega^2 = 2$, i.e.\ an additional $\Zed_2$ grading.
In particular, the vacuum occurs with multiplicity 2 in the fusion
of a $R$ field with itself indicating that all $R$ fields
correspond to fixed points in the fusion algebra of the
bosonic projection.
The structure of the fusion algebra is evident because the two matrices
$S^{NS,NS}$ and $S^{\NSS,R}$ are, up to a constant, equal
to the $S$-matrix of the Virasoro minimal $c(2,p)$ model.
The element $\omega$ of order 2 leads
to the correct sector structure due to
$\Zed_2$ charge conservation. Note that a rescaling of the
fields in the $R$ sector with a factor $1\over\sqrt{2}$ leads to a
fusion algebra where all coefficients are equal to zero or one and
the fusion charge conjugation is equal to $S^2$.
However, this rescaling is unnatural as our considerations
in section 3 have shown.
\sn
Note that the fusion algebras of these $\w(2,\delta)$-algebras
were also determined in ref.\ $\q{\roes}$ using
path space realizations of the corresponding characters.
\bn
Let us now discuss the \algs $\w(2,{{p-4}\over2})$ existing
for the Virasoro minimal values $c(p,12)$ with
$p\ge 5$ odd and $p,3$ coprime. The partition function
is of the type $(E_6,A_{p-1})$. From table $1$ we find that the
$\w$ characters are given by $\q{\blm,\wirrep}$:
$$\eqalign{
NS: \ \ \ \chi^{\w,NS}_i &= \chi_{1,i}+\chi_{7,i}+
                      \chi_{5,i}+\chi_{11,i} \ \ \ \ \ \ \
\ \ \ \ \ \ \ \ \ \ \ \ \ \ \,
     {\rm for} \ \ i \in {\cal I}_{NS} =
               \{1,...,{\scriptstyle{p-1\over2}} \} \cr
\NSS: \ \ \ \chi^{\w,\NSS}_i &= {\rm sgn}(p-4i)
\bigl(\chi_{1,i}+\chi_{7,i}-\chi_{5,i}-\chi_{11,i}\bigr)\ \ \ \
     {\rm for} \ \ i \in {\cal I}_{NS} =
               \{1,...,{\scriptstyle{p-1\over2}} \} \cr
R: \ \ \ \ \chi^{\w,R}_i \ &= 2(\chi_{4,i}+\chi_{8,i})
\,\ \ \ \ \ \ \ \ \ \ \ \ \ \ \ \ \ \ \ \ \ \ \ \ \ \ \ \ \ \ \ \ \ \ \
     {\rm for} \ \ i \in {\cal I}_{R} =
                \{1,...,{\scriptstyle{p-1\over2}} \}.
}\eqno{(4.5)}$$
Using the $S$-matrix of the minimal models of
the \vir we obtain for the $S$-matrix in the $\w$-character basis:
$$\eqalign{
S^{NS,NS}_{i,j} &= {2\over{\sqrt{p}}}
     (-1)^{i+j+1+\lfloor {p\over{12}} \rfloor}
{\rm sin}\bigl({12\pi i j \over p}\bigr)
\ \ \ \ \ \ \ \ \ \ \ \ \ \ \
\ \ \ \ \ \ \ \ \ \ \ \ \ \ \ \ \ \ \
                   {\rm{for}} \ \ i,j \in {\cal I}_{NS} \cr
S^{\NSS,R}_{i,l} \ \ &= {1\over2}S^{R,\NSS}_{l,i}
 =  {2\over{\sqrt{2p}}}{\rm sgn}(p-4i)(-1)^{l+(p\ {\rm mod}\ 3)}
  {\rm sin}\bigl({12\pi i l \over p}\bigr) \ \ \ \
  {\rm{for}} \ \ i \in {\cal I}_{NS},\ l \in {\cal I}_{R}
}\eqno{(4.6)}$$
The $S$-matrix obeys $(2.8)$ with
$H^R = (D^R)^{-1}={\textstyle{1\over2}}\id$.
As in the first example the resulting fusion algebra is isomorphic to the
fusion \alg of the Virasoro $(2,p)$ model tensor an element of
order two generating the right sector structure. This can be inferred from
the form $(4.6)$ of the $S$-matrix being equal to that of the $(2,p)$ model
modulo factors.
\mn
We remark that for $c(p,12)$ also the bosonic $\w$-\algs $\w(2,p-3)$ exist
and diagonalize the modular invariant partition functions given by
$$Z_{bos} =\sum\limits^{\scpmoneh}_{\scso}\bigl(
          {\scriptstyle{\mid\chi_{1,s}+\chi_{7,s}\mid}^2 +
                       {\mid\chi_{5,s}+\chi_{11,s}\mid}^2 +
                       {\mid\chi_{4,s}+\chi_{8,s}\mid}^2}\bigr)
\eqno{(4.7)}$$
The first realization for low spins is $\w(2,8)\subset\w(2,\sh)$ at
$c={21\over{22}}$ $\q{\blm}$. The subalgebra $\w(2,p-3)$ is the bosonic
projection of the fermionic \alg $\w(2,{p-4\over2})$ for $c=c(12,p)$
with $p\ge 5$ odd and $p,3$ coprime. One can show that
the field with conformal dimension ${p-4}\over2$ is a `simple current'
in the fusion algebra of the bosonic algebra. It is not very hard to find
the fusion algebra of the fermionic algebra from that of the bosonic
subalgebra (which can be calculated with the ordinary Verlinde formula) using
a `simple current' argument. In the appendix
we discuss this in detail for the $N=1$ super Virasoro algebra.
For a detailed discussion of `simple currents' in the context of
modular invariants of RCFTs we refer the reader to ref.\ $\q{\schell}$.
\bn
In the two series discussed above it was possible to obtain
a fusion \alg with fusion coefficients $\n_{ii}^1=1$ by rescaling the
$R$ fields with a factor $1\over{\sqrt{2}}$. In the general
case of the $(A,D)$ series this rescaling
would result in irrational fusion coefficients.
Already the simplest nontrivial example demonstrates this: We show that
the coefficients $\n_{ii}^1 = 2 \ ({\rm{for\ some\ }}i\in{\cal I}_R)$
are essential and cannot be removed by an extension
of the fusion algebra.
\mn
Our example is $\w(2,\dh)$ at $c=c(3,8)=-{21\over4}$ with the modular
invariant $(D_5,A_2)$. The model is the non-unitary super Virasoro
minimal model $c(2,8)$ and consists of two HWRs in each sector:
$$\vbox{\settabs 2 \columns
\+ $NS:\ \{0,{\textstyle{-{1\over4}}}\}\equiv\{\id,\sigma\}$ &
   $R:\ \{{\textstyle{-{3\over{32}},-{7\over{32}}}}\}\equiv\{\phi,\psi\}.$\cr
}$$
The $\w$-characters in terms of Virasoro characters are
$$\vbox{
\settabs\+\indent&$\chi^{\w,NS}_0=\chi_{1,1}+\chi_{1,2}$\hskip 1cm&\cr
\+ & $\chi^{\w,NS}_0 = \chi_{1,1} + \chi_{1,2}$ &
     $\chi^{\w,NS}_{-{1\over4}}\ \ = \chi_{3,1} + \chi_{3,2}$\cr
\+ & $\chi^{\w,R}_{-{3\over{32}}} = 2\chi_{4,1}$ &
     $\chi^{\w,R}_{-{7\over{32}}}\ \ = \chi_{2,1} + \chi_{2,2}\,.$\cr}$$
The matrix $D^R$ is equal to $diag(\{2,1\})$.
The fusion rules are given by (using the formulae of section 2):
$$\vbox{\settabs 3 \columns
\+ $\q{\sigma}\q{\sigma} = \q{\id} + 2\q{\sigma}$ &
   $\q{\sigma}\q{\phi} = \q{\phi} + 2\q{\psi}$    &
   $\q{\sigma}\q{\psi} = \q{\phi} + \q{\psi}$    \cr
\+ $\q{\phi}\q{\phi} = 2\q{\id} + 2\q{\sigma}$    &
   $\q{\phi}\q{\psi} = 2\q{\sigma}$           &
   $\q{\psi}\q{\psi} = \q{\id} + \q{\sigma}$      \cr
}$$
We have indeed $\n_{ii}^1 = D^R_{i,i}\ (i\in{\cal I}_R)$ but
a simple rescaling of the $R$ characters does not remove
the coefficient $\n_{\phi\phi}^1=2$.
Let us now show that an extension of this fusion \alg by splitting the
field $\phi$ is not possible. This is in contrast to
the case of degeneracies.
To this end one makes the most general ansatz
$(a,b,a^{\pm},b^{\pm},d,e,f^{\pm},g^{\pm},h^{\pm}\in\BN_0)$:
$$\vbox{\settabs\+
$\q{\phi^+}\q{\sigma}\!=\!f^+\q{\phi^+}\!+\!f^-\q{\phi^-}\!+\!h^+\q{\psi}$
\hskip 0.17cm &
$\q{\phi^-}\q{\sigma}\!=\!g^+\q{\phi^+}\!+\!g^-\q{\phi^-}\!+\!h^-\q{\psi}$
\hskip 0.17cm & \cr
\+ $\q{\phi} = \q{\phi^+} + \q{\phi^-}$
 & $\q{\phi^+}\q{\psi} = a \q{\sigma}$
 & $\q{\phi^-}\q{\psi} = b \q{\sigma}$\cr
\+ $\q{\phi^+}\q{\phi^+} = a^+\q{\id} + b^+\q{\sigma}$
 & $\q{\phi^-}\q{\phi^-} = a^-\q{\id} + b^-\q{\sigma}$
 & $\q{\phi^+}\q{\phi^-}\!=\!d\q{\id}\!+\!e\q{\sigma}$\cr
\+ $\q{\phi^+}\q{\sigma}\!=\!f^+\q{\phi^+}\!+\!f^-\q{\phi^-}\!+\!h^+\q{\psi}$
 & $\q{\phi^-}\q{\sigma}\!=\!g^+\q{\phi^+}\!+\!g^-\q{\phi^-}\!+\!h^-\q{\psi}$
&\cr
}$$
Using the fusion rules for $\q{\phi}\q{\psi},\ \q{\phi}\q{\phi}$ and
$\q{\sigma}\q{\phi}$ we get from this ansatz the following equations:
$$\vbox{\settabs 3 \columns
\+ $a + b = 2$     & $a^+ + a^- + 2d = 2$ & $b^+ + b^- + 2e = 2$ \cr
\+ $f^+ + g^+ = 1$ & $f^- + g^- = 1$      & $h^+ + h^- = 2$      \cr
}$$
{}From the associativity of
$\q{\psi}\q{\phi^+}\q{\phi^+}$ and
$\q{\psi}\q{\phi^-}\q{\phi^-}$ we have
$$\vbox{\settabs\+$a f^+ = b^+$\hskip 0.5cm&$a f^- = b^+$ \hskip 0.5cm&
$a h^+ = a^+ + b^+$\hskip 0.5cm&$b g^+ = b^-$\hskip 0.5cm&
$b g^- = b^-$\hskip 0.5cm& \cr
\+ $a f^+ = b^+$        &  $a f^- = b^+$ &
   $a h^+ = a^+ + b^+$  &  $b g^+ = b^-$ &
   $b g^- = b^-$        & $b h^- = a^- + b^-$ \cr
}$$
and from $\q{\psi}\q{\phi^+}\q{\phi^-}$ one obtains
$$\vbox{\settabs\+$a g^+ = d$\hskip 0.5cm&$a g^- = d$\hskip 0.5cm&
$a h^- = c + d$\hskip 0.5cm&$b f^+ = d$\hskip 0.5cm&
$b f^- = d$\hskip 0.5cm& \cr
\+ $a g^+ = e$  & $a g^- = e$ & $a h^- = d + e$ &
   $b f^+ = e$  & $b f^- = e$ & $b h^+ = d + e$ \cr
}$$
Because of the triviality of the solutions $a=0$ or $b=0$
(in the sense that one recovers the original fusion algebra),
we conclude from $a+b=2$ that $a = b = 1$. Inserting this into
$a g^+ + b f^+ = 2 e$, we get the contradiction
$1 = f^+ + g^+ = 2 e$ (q.e.d.).
\mn
Due to these facts
one is forced to weaken the axioms of fusion algebras for fermionic
theories, i.e.\ one has to admit more general fusion charge
conjugation matrices.
\bn\sn
\section{5. $N=1$ super Virasoro minimal models}
\bn
For the $N=1$ super Virasoro minimal models the central charge,
the conformal dimensions of the HWRs and the characters
are given by (for the unitary case see ref.\ $\q{\goddard}$)
$$\eqalign{ c &= c(p,q) = {\textstyle{\dh}}
\bigl(1-2{{(p-q)^2}\over{pq}}\bigr)
\phantom{xxx}p,q\in\BN,\ (p\mid q)=1,\ p+q\in 2\BN\ \ \ {\rm or}\cr
\phantom{c}&\phantom{=c(p,q)={\textstyle{\dh}}(1-2{{(p-q)^2}\over{pq}})xxx}
\,p,q\in 2\BN,\ (\scph\mid\scqh)=1,\ \scph+\scqh\not\in 2\BN\cr
h(r,s) &= {{(pr-qs)^2-(p-q)^2}\over{8pq}}+{{1-(-1)^{r+s}}\over{32}}
\phantom{xxx}1\le r\le q-1,\ \!1\le s\le p-1\cr
\chi^{SVir,NS}_{r,s}(\tau) &= {e^{-2\pi i {1\over48}}
               \eta({\tau+1\over2})\over \eta(\tau)^2}
            \bigl(\theta_{pr-qs,pq}({\tauh}) -
             \theta_{pr+qs,pq}({\tauh})\bigr)\phantom{xxxxxxxxx} \cr
\hat\chi^{SVir,R }_{r,s}(\tau) &=  {\eta(2\tau)\over \eta(\tau)^2}
                     \bigl(2-\del{r}{{q\over2}}\del{s}{{p\over2}}\bigr)
                     \bigl(\theta_{pr-qs,pq}({\tauh}) -
                    \theta_{pr+qs,pq}({\tauh})\bigr)
}\eqno{(5.1)}$$
\mn
where $r+s$ even (odd) corresponds to the $NS$ ($R$) sector. Note that we
do not include a `global' factor $\sqrt{2}$ in the $R$ characters
as in ref.\ $\q{\capsup}$ but use definition $(2.2)$.
Using the reflection symmetry in the superconformal
grid the set of linear independent characters is labelled by:
$$\eqalign{
NS:\ {\cal I}_{NS} &= \{\ (r,s)\ \vert \
r+s\ {\rm even},\ 1\le r\le q-1,\,1\le s\le\lfloor\scpmoneh\rfloor
\ {\rm or}\ 1\le r \le \scqh,\,s=\scph\}\cr
R:\ \ \ {\cal I}_{R} &= \{\ (r,s)\ \vert \
r+s\ {\rm odd},\ \ 1\le r\le q-1,\,1\le s\le\lfloor\scpmoneh\rfloor
\ {\rm or}\ 1\le r\le \scqh,\,s=\scph \}.
}\eqno{(5.2)}
$$
\mn
Using the well-known transformation properties of the theta functions under
modular transformations we obtain with a straightforward calculation
the following expressions for the $S$-matrix:
$$\eqalign{
S^{NS,NS}_{r_1,s_1;r_2,s_2}  &= {2\over\sqrt{pq}}
\bigl({\rm cos}\bigl({{2\pi\lambda_1\lambda_2}\over{4pq}}\bigr) -
{\rm cos}\bigl({{2\pi\bar\lambda_1\lambda_2}\over{4pq}}\bigr)\bigr)\cr
S^{R,\NSS}_{r_1,s_1;r_2,s_2} &= {2\over\sqrt{2pq}}
\bigl({\rm cos}\bigl({{2\pi\lambda_1\lambda_2}\over{4pq}}\bigr) -
(-1)^{r_2s_2}{\rm cos}\bigl({{2\pi\lambda_1\bar\lambda_2}\over{4pq}}
\bigr)\bigr)\cr
S^{\NSS,R}_{r_1,s_1;r_2,s_2} &= {2\over\sqrt{2pq}}
\bigl(1+\del{r_2}{{q\over2}}\del{s_2}{{p\over2}}\bigr)
\bigl({\rm cos}\bigl({{2\pi\lambda_1\lambda_2}\over{4pq}}\bigr) -
{\rm cos}\bigl({{2\pi\lambda_1\bar\lambda_2}\over{4pq}}\bigr)\bigr)\cr}
\eqno{(5.3)}$$
with $\lambda_i = pr_i-qs_i, \bar\lambda_i=pr_i+qs_i$.
One can check that with the standard definition
of $T$ these formulae define a proper \rep of the modular group.
It is well known that the {\it unitary} minimal models are
given by $(p,q)=(m,m+2)$ for $m\ge 2$. For this special choice of
$(p,q)$ formula $(5.3)$ reduces to the corresponding one already
given in $\q{\matsuo}$.
\mn
We emphasize that the multiplicity matrix $M$
is equal to the identity matrix in the case of the $N=1$
super Virasoro minimal models because there are
no additional independent `quantum numbers'.
\bn
Let us first consider the case $p$ and $q$ odd.
Here the reflection symmetry in the superconformal grid has no fixed
point and the $D$-matrix is given by $D^R=2\id$ since
the $S$-matrix obeys $(2.8)$ with $H^R = {\textstyle{1\over2}}\id$.
In order to calculate the fusion algebra corresponding to
the super Virasoro minimal models one has to insert $(5.3)$ into
$(2.5)$. It is not necessary to calculate the fusion
coefficients directly if one remembers that the
matrices $S^{NS,NS}$ and $S^{\NSS,R}$ are (modulo constants) equal
to the $S$-matrix of the Virasoro minimal models with
central charge $c=c(p,q)$, so that the well-known
selection rules for the Virasoro minimal models can be applied
to the fusion coefficients of the super Virasoro minimal models.
Since the selection rules of the Virasoro minimal models
respect the sector stucture given by odd or even $r+s$ in the
same way as the super minimal models, the corresponding
fusion algebras are isomorphic. Obviously, the only difference
between the fusion algebras is the
fact that the fusion coefficients connecting two $R$ fields
with a $NS$ field are elements of $2\BN$
for the $N=1$ supersymmetric model. Note that the fusion
algebra has a $\Zed_2$-structure like the first two examples
in section 4.
As in the case of the $\w(2,{k-2\over2})$ algebras,
$\n_{ii}^1 = 2 \ (\forall i\in {\cal I}_R)$
indicates that the fields in the $R$ sector
correspond to fixed points in the fusion algebra of the
bosonic projection.  Furthermore, the dimensions of the \reps of the
horizontal subalgebra in the $L_0$ eigenspace
${\cal V}_0$ are equal to the corresponding diagonal entries
of $D^R$.
\bn
In the case $p$ and $q$ even the structure is different.
Because $M$ is trivial the matrix $D$ can be immediately read
off from $(5.3)$ and $(2.8)$:
$D^R_{i,j} = (2-\delta_{i,i_0})\delta_{i,j}$, where $i_0$
is the label of the HWR $h={c\over{24}}$ in the $R$ sector.
Note that this is exactly the only fixed point under the
reflection symmetry in the superconformal grid:
$i_0\equiv(r,s) = (\scqh,\scph)$. This implies that
in this \rep $G_0^2$ is represented by zero so
that the irreducible \rep of the
horizontal subalgebra in ${\cal V}_0$ is one-dimensional
(in contrast to the generic case $G_0^2 \ne 0$
where the irreducible representations are two-dimensional).
Consequently, the corresponding $\widetilde{R}$ character
is nontrivial and encodes the broken boson-fermion
symmetry of the ground state. Hence the Witten index
$\q{\Witten,\Kastor}$ of this HWR is nontrivial.
This shows that
the diagonal entries of $D^R$ are equal to the dimensions of
the spaces ${\cal V}_0$.
Calculating the fusion \alg
with $(2.5)$ yields an associative and commutative \alg
with nontrivial fusion charge conjugation
$\n_{ij}^1 = D^R_{i,j}$.
The example $c(2,8)$ was already treated in section 4.
\bn
Finally, we present as a second example
the fusion algebra of the unitary model
$m=4$ with $c(4,6)=1$ $\q{\mus}$
corresponding to the $N=2$ supersymmetric
point of the Ashkin-Teller model.
There are 4 fields in the Neveu-Schwarz (Ramond) sector:
$h\in \{0,{1\over16},1,{1\over6}\}$
($h\in \{{3\over8},{1\over24},{9\over16},{1\over16}\}$).
The fusion algebra reads:
\sn
\def\nso{\cfs{NS}{0}}\def\nst{\cfs{NS}{{1\over{16}}}}
\def\nstr{\cfs{NS}{1}}\def\nsf{\cfs{NS}{{1\over6}}}
\def\ro{\cfs{R}{{3\over8}}}\def\rt{\cfs{R}{{1\over{24}}}}
\def\rtr{\cfs{R}{{9\over{16}}}}\def\rf{\cfs{R}{{1\over{16}}}}
$$\vbox{\settabs 2 \columns
\+ $\nst\nst = \nso + \nstr + 2\nsf$ & $\nst\nstr = \nst$               \cr
\+ $\nst\nsf = 2\nst$                & $\nstr\nstr = \nso$              \cr
\+ $\nstr\nsf = \nsf$                & $\nsf\nsf = \nso + \nstr + \nsf$ \cr
}$$
$$\vbox{\settabs 2 \columns
\+ $\ro\ro = 2\nso + 2\nstr$         & $\ro\rt = 2\nsf$                 \cr
\+ $\ro\rtr = 2\nst$                 & $\ro\rf = 2\nst$                 \cr
\+ $\rt\rt = \nso + \nstr + \nsf$    & $\rt\rtr = 2\nst$                \cr
\+ $\rt\rf = 2\nst$                  & $\rtr\rtr = 2\nso + 2\nsf$       \cr
\+ $\rtr\rf = 2\nstr + 2\nsf$        & $\rf\rf = 2\nso + 2\nsf.$        \cr
}\eqno{(5.4)}$$
\mn
The fusion rules for this model have also been calculated in $\q{\mus}$
using the Coulomb-gas approach. However, this approach shows only if
a fusion coefficient vanishes or not. Furthermore, null vector methods
do not work because the $G_0$-diagonal spin field is not
well-defined $\q{\mus}$.
\bn
Note that in $(5.4)$ a coefficient $2$ appears in front of the vacuum
representation in the fusion of all $R$ fields besides
$\Phi^{R}_{\scriptstyle{1\over{24}}}$
with itself. Usually, one demands that
the fusion of a field with its conjugate contains the vacuum
only once. However, one can check that the fusion
coefficients in $(5.4)$ are sufficient and necessary for the
associativity of the fusion algebra.
In contrast to the $(odd,odd)$ case discussed above, here
it is impossible to rescale the fields
in the $R$ sector in such a way that the resulting
fusion algebra is integer-valued and has a trivial fusion charge
conjugation. As was shown in section 3,
the choice of the normalization of the fields
in the $R$ sector is fixed if one requires
that the fusion algebra of the bosonic projection
of the $N=1$ supersymmetric algebra under consideration induces
the supersymmetric fusion algebra in a consistent way.
\bn
The fusion algebra $(5.4)$ contains a `simple current'
of conformal dimension $1$ and a subalgebra generated
from the fields $\nso,\nstr,\nsf,\ro,\rt$.
This is reminiscent of the additional $N=2$ supersymmetry of
the $c=1$ model. Taking into account the fact that the $N=2$
super Virasoro algebra has a minimal model at $c=1$ it is
obvious to consider the extension of the symmetry algebra by this
`simple current'.
\bn\mn
\section{6. $N=1$ $\sw(\dh,\delta)$-algebras}
\bn
In this section we investigate $N=1$ super $\w$-algebras
$\sw(\dh,\delta)$ with two generators for $\delta\ge2$. These
are the most complicated examples of fermionic RCFTs
as far as the fusion algebra is concerned
because both the multiplicity matrix $M$
as well as the $D$-matrix are different from the identity matrix
in the general case. After some general comments on the
representation theory in the Ramond sector we discuss
the $\sw(\dh,\delta)$-algebras fitting
into the $ADE$-classification $\q{\capsup}$.
Then we proceed with the parabolic
$N=1$ $\sw$-algebras with vanishing and non-vanishing self-coupling
constant. We give the matrices $M$ and $D$ for all these series.
As an example for $ADE$-cases we present the fusion
algebra for the rational model of $\sw(\dh,2)$ at $c=-{6\over5}$.
We conclude with some further remarks concerning fusion algebras of
fermionic RCFTs possessing a $\sw(\dh,\delta)$ symmetry algebra.
\bn
We found by explicit computer calculations that for the
$\sw(\dh,\delta)$-algebras with
$2\le \delta \le {9\over2}$ the identity
$\lb \psi_0,G_0\phi_0 \rb = 0$
holds on the corresponding highest weight vectors
where $\psi$ $(\phi)$ is the bosonic (fermionic) component of the
additional super field. Therefore, one considers only HWRs in
which $G_0\phi_0$ is represented by a scalar on the highest weight
leading to an additional quantum number which can take at most two
values (since $G_0\phi_0$ satisfies a quadratic equation for
fixed $L_0$ and $\psi_0$ eigenvalues of the highest weight)
(cf. section 2). However, this quantum number is redundant in the
supersymmetric theory because ${\cal V}_0$ contains for all
possible eigenvalues of $G_0\phi_0$ an eigenvector.
Nevertheless, in the bosonic projection it distinguishes
between different highest weight representations.
Because  $G_0\phi_0$ is represented by a scalar on the highest
weight the dimension of ${\cal V}_0$ is at most two, it is
one-dimensional exactly if $G_0^2 = \phi_0^2 = 0$ holds on the highest
weight. Indeed, for $\delta\in\BN$ the identity $G_0^2 =0$
implies $\phi_0^2 = 0$ due to $\lb G_0,\psi_0\rb = \phi_0$.
However, computer results show that even for
$\nh\ge\delta\in\BN+{1\over2}$ this implication is true.
In the case $\delta\in\BN+{1\over2}$, $G_0^2\not=0$
the representation of the horizontal subalgebra in ${\cal V}_0$
is equal to the two-dimensional representation
of the Clifford algebra $Cl(2,0)$
(cf. section 4). For $\delta\in\BN$ the structure of the
representation of the horizontal subalgebra is more complicated.
\bn
The algebras existing for super Virasoro minimal values of $c$
can be organized into four series according to the partition
function which is diagonalized by the $\sw$-characters. From
table $2$ we can directly read off these
characters in terms of super Virasoro characters (the quantities
appearing with their absolute value in $Z$ are the $\sw$-characters).
Obviously, one can obtain $D$ from the form of the
partition functions given in table $2$ if the multiplicity
matix $M$ is known. This matrix follows from earlier
studies of the degeneracies of the \rep
modules $\q{\eh1,\ralfdip}$. The representations whose characters
are a sum of Virasoro characters with maximal number
of summands are non-degenerate.
The representations whose characters are a sum
of Virasoro characters with half the number of summands are doubly
degenerate. Note that one has to consider the $NS$ and $R$ sectors
separately.
\sn
\vbox to 5.5cm{}
\mn
\centerline{\vbox{
\hbox{\vbox{\offinterlineskip
\def\tablespace{ height2pt&\omit&&\omit&&\omit&&\omit&\cr }
\def\smtablespace{ height0.5pt&\omit&&\omit&&\omit&&\omit&\cr }
\def\btablespace{ height4pt&\omit&&\omit&&\omit&&\omit&\cr }

\def\tablerule{ \tablespace\noalign{\hrule}\tablespace}
\hrule\halign{&\vrule#&\strut\hskip0.05cm\hfil#\hfil\hskip0.1cm\cr
\tablespace
& $\sw(\dh,\delta)$  && $c(p,q)$  && series && $Z^{NS},Z^R$  &\cr
\tablerule\tablerule
& ${{k-2}\over2}$  && $c(12,2k)$ && $(D_{k+1},E_6)$ &&
$Z^{NS} =
\sum\limits^{k-1}_{{r=1}\atop{odd}}
{\scriptstyle\mid\chi_{r,1}+\chi_{r,7}+\chi_{r,5}+\chi_{r,11}\mid^2}+$ &\cr
\smtablespace
& \omit  && $k$ odd && \omit &&
\phantom{xxxxxxxxxxxxxx}
${\scriptstyle 2\mid\chi_{k,1}+\chi_{k,5}\mid^2}$ &\cr
\tablespace
& \omit  && $(6,k)=1$ && \omit &&
$Z^R = \sum\limits^{k-1}_{{r=1}\atop{odd}}
{\scriptstyle \mid \hat\chi_{r,4}+\hat\chi_{r,8}\mid^2} +
{\scriptstyle 2\mid \hat\chi_{k,4}\mid^2}$ &\cr
\tablerule
& ${{2k-3}\over2}$  && $c(12,2k)$ && $(A_{2k-1},E_6)$ &&
$Z^{NS}\!=
\!\sum\limits^{k-1}_{{r=1}\atop{odd}}\!\bigl(
{\scriptstyle\mid\chi_{r,1}+\chi_{r,7}\mid^2+
             \mid\chi_{r,5}+\chi_{r,11}\mid^2}\bigr)+$ &\cr
\smtablespace
& \omit  && $k$ odd && \omit &&
\phantom{xxxx}
${\scriptstyle\mid\chi_{k,1}+\chi_{k,7}\mid^2}+
\sum\limits^{k-1}_{{r=2}\atop{even}}\!
{\scriptstyle\mid\chi_{r,4}+\chi_{r,8}\mid^2}$ &\cr
\tablespace
& \omit  && $(6,k)=1$ && \omit &&
$Z^R\!=\!\sum\limits^{k-1}_{{r=2}\atop{even}}\!\bigl(
{\scriptstyle {1\over2}\mid \hat\chi_{r,1}+\hat\chi_{r,7}\mid^2 +
{1\over2}\mid \hat\chi_{r,5}+\hat\chi_{r,11}\mid^2}\bigr)+$&\cr
\smtablespace
& \omit  && \omit && \omit &&
\phantom{xxxxxxxx}
$\sum\limits^{k-1}_{{r=1}\atop{odd}}\!
{\scriptstyle {1\over2}\mid \hat\chi_{r,4}+\hat\chi_{r,8}\mid^2}\!+\!
{\scriptstyle \mid \hat\chi_{k,4}\mid^2}$ &\cr
\tablerule
& ${{2k-5}\over2}$ && $c(30,2k)$ && $(A_{2k-1},E_8)$ &&
$Z^{NS} =\sum\limits^{k-1}_{{r=1}\atop{odd}}\!\bigl(
{\scriptstyle\mid\chi_{r,1}+\chi_{r,11}+\chi_{r,19}+\chi_{r,29}\mid^2}+$ &\cr
\smtablespace
& \omit  && $k$ even && \omit &&
\phantom{xxxxxxxxxxxxxx}
${\scriptstyle\mid\chi_{r,7}+\chi_{r,13}+\chi_{r,17}+
\chi_{r,23}\mid^2}\bigr)$ &\cr
\tablespace
& \omit  && $(15,k)=1$ && \omit &&
$Z^R\!=\!\sum\limits^{k-1}_{{r=2}\atop{even}}\!\bigl(
{\scriptstyle {1\over2}\mid \hat\chi_{r,1}+\hat\chi_{r,11}+
\hat\chi_{r,19}+\hat\chi_{r,29}\mid^2}+$ &\cr
& \omit  && \omit && \omit &&
\phantom{xxxxxxxxxxxx}
${\scriptstyle {1\over2}\mid \hat\chi_{r,7}+\hat\chi_{r,13}+
\hat\chi_{r,17}+\hat\chi_{r,23}\mid^2}\bigr)+$ &\cr
\btablespace
& \omit  && \omit && \omit &&
\phantom{xxxxxxxxxxxx}
${\scriptstyle \mid \hat\chi_{k,1}+
\hat\chi_{k,11}\mid^2}\!+\!
{\scriptstyle \mid \hat\chi_{k,7}+
\hat\chi_{k,13}\mid^2}$ &\cr
\tablerule
& ${{(q-2)(k-1)}\over4}$ && $c(2k,q)$ && $(A_{q-1},D_{k+1})$ &&
$Z^{NS} =\sum\limits^{{q\over2}-1}_{{r=1}\atop{odd}}\!\Bigl(
\sum\limits^{k-1}_{{s=1}\atop{odd}}
{\scriptstyle\mid\chi_{r,s}+\chi_{q-r,s}\mid^2}+
{\scriptstyle 2\mid\chi_{r,k}\mid^2}\Bigr)$ &\cr
\tablespace
& \omit  && $k$ odd && \omit &&
$Z^R\!=\!\sum\limits^{{q\over2}-1}_{{r=2}\atop{even}}
\sum\limits^{k-1}_{{s=1}\atop{odd}}
{\scriptstyle {1\over2}\mid \hat\chi_{r,s}+\hat\chi_{q-r,s}\mid^2}+
\sum\limits^{k-1}_{{s=1}\atop{odd}}
{\scriptstyle \mid \hat\chi_{{q\over2},s}\mid^2}+$ &\cr
\smtablespace
& \omit  && $(k,{q\over2})=1$ && \omit &&
\phantom{xxxxxxx}
$\sum\limits^{{q\over2}-1}_{{r=2}\atop{even}}
{\scriptstyle \mid \hat\chi_{r,k}\mid^2}+
{\scriptstyle 2\mid\hat\chi_{{q\over2},k}\mid^2}$ &\cr
\tablespace}\hrule}}
\hbox{\hskip 0.5cm Table 2: partition functions and series
of $\sw(\dh,\delta)$-algebras $\q{\cap,\blmwir,\eh1}$}
}}
\mn
We continue with the discussion of parabolic $\sw$-algebras. There
are two series, one with non-vanishing and one with vanishing
self-coupling constant $\q{\pim}$. The series with
$C_{\phi\phi}^{\phi}\not=0$ consists of the \algs $\sw(\dh,8k)$ at
$c={\textstyle{3\over2}}(1-16k)$ with $4k\in\BN$. They possess the
following HWRs $\q{\eh1,\ralfdip}$
($h_{r,r}=k(r^2-1),\ h_{r,-r}=h_{r,r}+{1\over2}r^2$):
\sn
\settabs\+\indent&$NS$\hskip 1cm &$h_{{2m+1\over{4k+4}},
-{2m+1\over{4k+4}}}$xxxxx\hskip 1cm&\cr
\+&$NS:$&$h_{{m\over{4k}},{m\over{4k}}}$&$m = 0,\dots,4k,8k$\cr
\+&     &$h_{{m\over{4k+2}},-{m\over{4k+2}}}$ &$m=1,\dots,4k+1$\cr
\+&$R:$ &$h_{{m\over{4k}},{m\over{4k}}}+{\textstyle{1\over{16}}}$
&$m = 0,\dots,4k$\cr
\+& &$h_{{m\over{4k+2}},-{m\over{4k+2}}}
+{\textstyle{1\over{16}}}$ &$m=0,\dots,4k+2$\cr\noindent
\sn
The modular invariant partition function $Z$
is given by the expressions $\q{\pim}$:
$$\eqalign{
Z^{NS}(\tau) &=
\biggl\vert{\eta({\scriptstyle{\tau+1}\over2})\over{\eta(\tau)^2}}
\biggr\vert^2\Bigl(
\mid{\textstyle{1\over2}}(\theta_{4k,4k}(\tau)-
\theta_{4k,4k+2}(\tau))\mid^2 +
\mid{\textstyle{1\over2}}(\theta_{4k,4k}(\tau)+
\theta_{4k,4k+2}(\tau))\mid^2 +\cr
\phantom{Z^{NS}(\tau)} &\phantom{=
\biggl\vert{\eta({\scriptstyle{\tau+1}\over2})\over{\eta(\tau)^2}}
\biggr\vert^2\Bigl(}
\mid{\textstyle{1\over2}}(\theta_{0,4k}(\tau)-
\theta_{0,4k+2}(\tau))\mid^2 +
\mid{\textstyle{1\over2}}(\theta_{0,4k}(\tau)+
\theta_{0,4k+2}(\tau))\mid^2 +\cr
\phantom{Z^{NS}(\tau)} &\phantom{=
\mid{\eta({\scriptstyle{\tau+1}\over2})\over{\eta(\tau)^2}}
\mid^2\Bigl(}
\sum_{m=1}^{4k+1} \mid\theta_{4k+2+m,4k+2}(\tau)\mid^2 +
\sum_{m=1}^{4k-1} \mid\theta_{4k+m,4k}(\tau)\mid^2\Bigr)\cr
Z^{R}(\tau) &=
\biggl\vert{\eta(2\tau)\over{\eta(\tau)^2}}\biggr\vert^2\Bigl(
\mid\theta_{0,4k+2}(\tau)\mid^2 + \mid\theta_{0,4k}(\tau)\mid^2 +
\mid\theta_{4k,4k}(\tau)\mid^2 + \mid\theta_{4k+2,4k+2}(\tau)\mid^2 +\cr
\phantom{Z^{R}(\tau)} &
\phantom{= \biggl\vert{\eta(2\tau)\over{\eta(\tau)^2}}\biggr\vert^2\Bigl(}
\sum_{m=1}^{4k-1}
{\textstyle{1\over2}}\mid 2\theta_{4k+m,4k}(\tau)\mid^2 +
\sum_{m=1}^{4k+1}
{\textstyle{1\over2}}\mid 2\theta_{4k+2+m,4k+2}(\tau)\mid^2\Bigr)\cr}
\eqno{(6.1)}$$
Firstly, we state that the {\it two} \rep modules to the $h$-value
$h_{0,0}+{1\over{16}} = {c\over{24}}$ $-$ corresponding to the first
two summands in $Z^R$ $-$ are different and that the dimension
of the \rep of the horizontal subalgebra in ${\cal V}_0$ is
equal to one in both cases.
In all other \reps this dimension is equal to two.
Furthermore, the \reps to $h_{1,1}+{1\over{16}}$
$-$ third summand $-$ and
$h_{1,-1}+{1\over{16}}$ $-$ fourth summand $-$
are doubly degenerate, whereas all
other \reps occur only once. We conclude that
$M^R=diag(\{1,1,2,2,1,\dots,1\})$ so that
$D^R=diag(\{1,1,2,2,2,\dots,2\})$ (again the diagonal entries of
$D^R$ equal the dimensions of the corresponding spaces ${\cal V}_0$).
\sn
The series with $C_{\phi\phi}^{\phi}=0$ is given
by the \algs $\sw(\dh,3k)$ existing
for $c={\textstyle{3\over2}}(1-16k)$ with $2k\in\BN$.
They possess the following HWRs $\q{\eh1,\ralfdip}$:
\sn
\settabs\+\indent&$NS$\hskip 1cm
&$h_{{2m+1\over{4k+4}},-{2m+1\over{4k+4}}}$xxxxx\hskip 1cm&\cr
\+&$NS:$&$h_{{m\over{2k}},{m\over{2k}}}$&
$m = 0,\dots,\lfloor k \rfloor,2k$\cr
\+&     &$h_{{m\over{2k+1}},-{m\over{2k+1}}}$ &
$m=0,\dots,\lfloor k + {\textstyle{1\over2}}\rfloor $\cr
\+&$R:$ &$h_{{m\over{2k}},{m\over{2k}}}+{\textstyle{1\over{16}}}$
&$m = k-\lfloor k \rfloor,\dots,k$\cr
\+& &$h_{{m\over{2k+1}},-{m\over{2k+1}}}
+{\textstyle{1\over{16}}}$ &
$m={\textstyle{1\over2}}-(k-\lfloor k \rfloor),\dots,
k+{\textstyle{1\over2}}$\cr\noindent
\sn
It turned out in the course of our calculations that
in the Ramond sector one has to distinguish the two cases
$k\in\BN$ and $k\in\BN+{1\over2}$.
In the case $k\in\BN$ there are two doubly degenerate HWRs with
two-dimensional $-$ $d=2$ $-$ \rep of the horizontal subalgebra
$(h_{{1\over2},{1\over2}}+{1\over{16}},
h_{{1\over2},-{1\over2}}+{1\over{16}})$.
In the single \rep with $h_{0,0}+{1\over{16}}={c\over{24}}$
this dimension $d$ is equal to one. All other
\reps are non-degenerate and have $d=2$.
For $k\in\BN+{1\over2}$ there is only one doubly degenerate \rep
$(h_{{1\over2},{1\over2}}+{1\over{16}})$ which has $d=2$.
Furthermore, the \rep $h_{0,0}+{1\over{16}}={c\over{24}}$
has $d=1$ and all other representations have $d=2$
and are not degenerate, clarifying some unexplained subtleties in
ref.\ $\q{\pim}$.
{}From the corresponding modular invariant partition function
$Z$ one obtains $D^R$.
For $k\in\BN$ the diagonal entries of $D^R$ are given by the
dimensions $d$ of ${\cal V}_0$. For $k\in\BN+{1\over2}$ this
is only different for the representation
with conformal dimension  $(h_{{1\over2},-{1\over2}}+{1\over{16}})$
where $d=2$ but
the $D^R$ entry is equal to one. Here we see that in general the entries
of the diagonal matrix $D^R$ are different from the corresponding
dimensions of ${\cal V}_0$.
\bn
We present now an example for a fusion \alg of a rational
model of a $\sw(\dh,\delta)$-algebra which shows the most
general features of fusion \algs of fermionic RCFTs.
Our example is $\sw(\dh,2)$ at $c=-{6\over5}$ corresponding
to the partition function $(A_3,D_6)$. The model consists of three
HWRs in each sector:
$NS:h\in\{0,-{1\over{10}},{1\over5}\}$,
\hskip 0.5cm
$R:h\in\{{3\over4},{3\over{20}},-{1\over{20}}\}$. The last
$h$-value is the fixed point of the superconformal grid and
hence equal to $c\over{24}$. The \reps with
$h={1\over5},{3\over4},{3\over{20}},-{1\over{20}}$ are doubly
degenerate. The dimensions of the \reps of the horizontal
subalgebra in ${\cal V}_0$ are equal to {\it two} with the exception of
the \rep $h=-{1\over{20}}={c\over{24}}$, where
it is equal to {\it one}. These dimensions coincide again with
the corresponding diagonal entries of $D^R$ obtained from $M$ and
$S$. Using the explicit form of the
$\sw$-characters in terms of super Virasoro characters, we obtain
the fusion algebra via formulae $(4.3)$ and $(2.5)$:
\def\nso{\cfs{NS}{0}}\def\nst{\cfs{NS}{-{1\over{10}}}}
\def\nstr{\cfs{NS}{{1\over5}}}
\def\ro{\cfs{R}{{3\over4}}}\def\rt{\cfs{R}{{3\over{20}}}}
\def\rtr{\cfs{R}{-{1\over{20}}}}
$$\vbox{\settabs 2 \columns
\+ $\nst\nst = \nso + \nst + \nstr$ & $\nst\nstr = 2\nst + \nstr$ \cr
\+ $\nstr\nstr = 2\nso + 2\nst + \nstr$ & \cr
\+ & \cr
\+ $\ro\ro = 4\nso$        & $\ro\rt = 4\nst$ \cr
\+ $\ro\rtr = 2\nstr$      & $\rt\rt = 4\nso + 4\nst + 4\nstr$ \cr
\+ $\rt\rtr = 4\nst + 2\nstr$&$\rtr\rtr = 2\nso + 2\nst + \nstr$ \cr
}\eqno{(6.2)}$$
\sn
We recognize that the equality $\n_{ii}^1=(MD)_{i,i}$ ($i\in{\cal I}_R$)
is indeed satisfied in this example. Using the fact that the
fusion algebra of the super Virasoro minimal model $c(10,4)=-{6\over5}$
possesses a $\Zed_2$ `simple current' of conformal dimension $2$
one can recover $(6.1)$ from the fusion algebra of the
super Virasoro minimal model.
\mn
Finally, note that it is possible to obtain the fusion algebras
of the $\sw(\dh,\delta)$-algebra rational models related to the
$(A_{q-1},D_{{p+2}\over2})$-series in the $ADE$-classification
from the super Virasoro fusion algebras using the simple current
of order two and conformal dimension $\delta$. However,
this is not possible for $\sw(\dh,\delta)$-algebras fitting into
one of the other three series because the field with
conformal dimension $\delta$ is no `simple current' in the
fusion algebra of the corresponding super Virasoro minimal model
any more. Nevertheless, it is possible to obtain the fusion
algebra with the generalized Verlinde formula from the
$S$-matrix of the $\sw(\dh,\delta)$-minimal model.
Furthermore, one can show that $-$ in perfect
analogy to the second example in section 4 $-$
the fusion algebras of the models related to $(D_{k+1},E_6)$
factorize into a $\Zed_2\otimes{\cal A}$ fusion algebra
(${\cal A}$ denotes the fusion algebra of the $NS$ sector).
In the first example $-$ $\sw(\dh,\fh)$ at $c=c(12,14)={10\over7}$ $-$
we verified by explicit calculation that it is
possible to resolve the degeneracies in the $NS$ sector
by a suitable extension of the fusion algebra.
\vskip 0pt\noindent
Unfortunately, we are not able to calculate the
fusion algebras of the exceptional rational models of
$\sw(\dh,\delta)$ algebras since no explicit formulae for the
$S$-matrices are known.
\bn
\vfill
\eject
\section{7. Conclusions }
\sn
We proved a generalized Verlinde formula for
fermionic RCFTs by showing that the fusion algebras
coincide with those obtained from the corresponding
bosonic projection by the ordinary Verlinde formula
and `simple current' arguments.
Using this generalized Verlinde formula we were able to
calculate the fusion algebras of several fermionic RCFTs.
The $S$-matrix is in general neither unitary nor symmetric
but it obeys the equation $S^{\dagger}HS=H$ with
$H=MD^{-1}$. $M$ is a diagonal matrix encoding the
multiplicities of the HWRs of the theory whereas
$D$ is a diagonal matrix which is defined through the
orbit lengths under the action of the `simple current' in
the fusion algebra of the bosonic projection (cf.\ section 2).
In a concrete example we showed that it is not possible
to avoid the $D$-matrix by extending the fusion algebra.
There is strong evidence that this holds in general.
Furthermore, we considered the representation theory
of the horizontal subalgebra on the highest weights in
the Ramond sector for fermionic $\w$-algebras.
For fermionic $\w(2,\delta)$-algebras the dimensions of
the irreducible representations are encoded in the fusion
algebra of the corresponding rational model.
In particular, we considered fermionic $\w(2,\delta)$-algebras,
minimal models of the $N=1$ super Virasoro
algebra and finally $N=1$ $\sw$-algebras with two generators.
In some cases we verified explicitly that our results agree with
the fusion algebras calculated by `simple current' arguments
from bosonic projections of the corresponding fermionic theories.
Furthermore, we pointed out that in the case of the $N=1$
super Virasoro minimal models with central charge $c = c(p,q)$,
$p,q$ even, null-state methods cannot be applied.
These examples show that one has to weaken the axioms of
fusion algebras for fermionic RCFTs allowing more general fusion
charge conjugation matrices $\n_{ij}^1=(MD)_{i,j}$.
We have shown that in general fusion coefficients greater than
one appear, so the consideration of 3-point functions
in the Coulomb-gas picture which only tells whether a
certain field appears in the fusion of two other fields or not,
cannot yield the full information about the fusion algebra.
\vskip 0pt\noindent
Using `simple current' arguments it is possible to define
in a very natural way the fusion of twisted fields in the
cases of bosonic $\w(2,\delta)$-algebras admitting an outer
$\Zed_2$-automorphism. It is not clear to us how to write
down a generalized Verlinde formula in these cases.
\vskip 0pt\noindent
It will be interesting to use the generalized Verlinde formula
in the future to set up a classification program for the fusion
algebras of fermionic theories in complete analogy to the bosonic case
considered in $\q{\cas,\ehf}$. Especially it would be very
interesting to show whether fermionic RCFTs exist which cannot
be obtained by a $\Zed_2$ `simple curent' extension of its bosonic
projection.
\mn
\section{Acknowledgements}
\sn
We are very grateful to W.\ Nahm for encouraging
us to study fusion algebras of bosonic projections and
many valuable discussions.
\vskip 0pt\noindent
We thank R.\ Blumenhagen, M.\ Flohr, A.\ Honecker,
J.\ Kellendonk, S.\ Mallwitz, N.\ Mohammedi,
A.\ Recknagel, M.\ R\"osgen, N.-P.\ Skoruppa,
M.\ Terhoeven and R.\ Varnhagen for useful discussions.
\vskip 0pt\noindent
W.E.\ thanks the Max-Planck-Institut f\"ur Mathematik for financial support.
\vskip 0pt\noindent
R.H.\ is supported by a research studentship
of the NRW-Graduiertenf\"orderung.
\bn
\vfill\eject
\section{Appendix}
\mn
In this appendix the formulae of section 3 are verified by
considering the bosonic projection of the $N=1$ super Virasoro
algebra. We calculate the fusion algebra $-$ using the ordinary
Verlinde formula $-$ of the bosonic projection for
two special minimal values of $c$
where this projection yields a $\w(2,4)$.
One recognizes that these fusion algebras contain a
`simple current' of dimension $\dh$.
The definition of a new basis, given by
the sum of the fields lying in one orbit, divided by the length
of the orbit, yields the fusion algebra of the $N=1$ super
Virasoro minimal model. The fusion algebras obtained this
way coincide exactly with the fusion algebras calculated
directly using the generalized Verlinde formula.
\bn
It is well-known that the projection of the $N=1$ \supvir onto the bosonic
sector yields a $\w(2,4,6)$ $\q{\bowknegt-\andicorr}$. As one can easily show
by direct computation, the primary field of conformal dimension
six turns out to be a null field for some special values of $c$, among them
$-11$ and $-{11\over{14}}$, so that the $\w(2,4,6)$ reduces to a $\w(2,4)$
(the primary field of dimension four has nonzero norm for
these values of $c$) $\q{\andicorr}$.
It has been found earlier that at these two values of $c$ rational
models of $\w(2,4)$ exist and the possible $h$-values are known
$\q{\wirrep}$.
\sn
Firstly let us determine the fusion \alg of the $N=1$ super
Virasoro minimal model at $c(2,12)=-11$ using
$(4.3)$ and $(2.5)$. There are three HWRs in each sector:
$NS:h\in\{0,-{1\over3},-{1\over2}\}$,
\hskip 0.5cm
$R:h\in\{-{1\over8},-{3\over8},-{11\over{24}}\}$.
Note that the last $h$-value
is the fixed point of the superconformal grid and is
equal to $c\over{24}$. The fusion \alg reads:
\def\nso{\cfs{NS}{0}}\def\nst{\cfs{NS}{-{1\over3}}}
\def\nstr{\cfs{NS}{-{1\over2}}}
\def\ro{\cfs{R}{-{1\over8}}}\def\rt{\cfs{R}{-{3\over{8}}}}
\def\rtr{\cfs{R}{-{11\over{24}}}}
$$\vbox{\settabs 2 \columns
\+ $\nst\nst = \nso + \nst + \nstr$     & $\nst\nstr = \nst + 2\nstr$ \cr
\+ $\nstr\nstr = \nso + 2\nst + 2\nstr$ & \cr
\+ & \cr
\+ $\ro\ro = 2\nso + 2\nst$             & $\ro\rt = 2\nst + 2\nstr$ \cr
\+ $\ro\rtr = 2\nstr$              & $\rt\rt = 2\nso + 2\nst + 4\nstr$ \cr
\+ $\rt\rtr = 2\nst + 2\nstr$      & $\rtr\rtr = \nso + \nst + \nstr$ \cr
}\eqno{(A.1)}$$
\sn
For $\w(2,4)$ at $c=-11$ the effective central charge $\tilde c$
is equal to $1$, so that this model belongs to the
parabolic $\w(2,\delta)$-algebras
which have been studied by M.\ Flohr $\q{\pim}$.
Using the explicit form of the $S$-matrix one is able
to calculate the fusion algebra of this model $\q{\pimpriv,\pim}$.
There exist $10$ HWRs of $\w(2,4)$ at $c=-11$:
\vskip 0.001pt\noindent
$h\in\{0,\dh,-{1\over3},{1\over6},-{1\over2},{\tilde 0},
-{1\over8},-{3\over8},-{11\over{24}},{13\over{24}}\}$.
The fusion \alg is given by:
\mn
\vbox to 1cm{}
\vfill
\eject
\def\bo{\cfs{}{0}}
\def\bt{\cfs{}{{3\over2}}}\def\btr{\cfs{}{-{1\over3}}}
\def\bfo{\cfs{}{{1\over6}}}\def\bfi{\cfs{}{-{1\over2}}}
\def\bsi{\cfs{}{\tilde 0}}\def\bse{\cfs{}{-{1\over8}}}
\def\ba{\cfs{}{-{3\over8}}}\def\bne{\cfs{}{-{11\over{24}}}}
\def\bz{\cfs{}{{13\over{24}}}}
$$\vbox{\settabs\+xx&$\bfi\bfi=\bo+\btr+\bfo+\bfi+\bsi$&\hskip 1cm\cr
\+& $\bt\bt = \bo$ & $\bt\btr = \bfo$ \cr
\+& $\bt\bfo = \btr$ & $\bt\bfi = \bsi$ \cr
\+& $\bt\bsi = \bfi$ & $\bt\bse = \bse$ \cr
\+& $\bt\ba = \ba$ & $\bt\bne = \bz$ \cr
\+& $\bt\bz = \bne$ & \cr
\+& & \cr
\+& $\btr\btr = \bo + \btr + \bfi$ & $\btr\bfo =\bt+\bfo+\bsi$ \cr
\+& $\btr\bfi = \btr +\bfi + \bsi$ & $\btr\bsi =\bfo+\bfi+\bsi$ \cr
\+& $\btr\bse = \bse + \ba$        & \cr
\+& $\btr\ba =\bse+\ba+\bne+\bz$    & \cr
\+& $\btr\bne = \ba + \bz$          & $\btr\bz = \ba + \bne$ \cr
\+& & \cr
\+& $\bfo\bfo = \bo + \btr + \bfi$ & $\bfo\bfi =\bfo+\bfi+\bsi$ \cr
\+& $\bfo\bsi = \btr + \bfi +\bsi$ & $\bfo\bse = \bse + \ba$ \cr
\+& $\bfo\ba = \bse +\ba+\bne+\bz$  & $\bfo\bne = \ba + \bne$ \cr
\+& $\bfo\bz = \ba + \bz$          & \cr
\+& & \cr
\+& $\bfi\bfi=\bo+\btr+\bfo+\bfi+\bsi$ & \cr
\+& $\bfi\bsi=\bt+\btr+\bfo+\bfi+\bsi$ & \cr
\+& $\bfi\bse = \ba + \bne + \bz$       & \cr
\+& $\bfi\ba =\bse+2\ba+\bne+\bz$       & \cr
\+& $\bfi\bne = \bse + \ba + \bne$       & $\bfi\bz = \bse + \ba + \bz$ \cr
}$$
$$\vbox{\settabs\+xx&$\bfi\bfi=\bo+\btr+\bfo+\bfi+\bsi$&\hskip 1cm\cr
\+& $\bsi\bsi=\bo+\btr+\bfo+\bfi+\bsi$ & $\bsi\bse = \ba + \bne + \bz$ \cr
\+& $\bsi\ba = \bse + 2\ba+\bne + \bz$  & \cr
\+& $\bsi\bne = \bse + \ba + \bz$       & $\bsi\bz = \bse + \ba + \bne$ \cr
\+ & & \cr
\+& $\bse\bse = \bo + \bt + \btr+\bfo$ & \cr
\+& $\bse\ba = \btr+\bfo+\bfi + \bsi$  & \cr
\+& $\bse\bne = \bfi + \bsi$            & $\bse\bz = \bfi + \bsi$ \cr
\+& & \cr
\+& $\ba\ba =\bo+\bt+\btr+\bfo+2\bfi+2\bsi$  & \cr
\+& $\ba\bne = \btr+\bfo+\bfi+\bsi$           & \cr
\+& $\ba\bz =\btr+\bfo+\bfi+\bsi$            & \cr
\+& & \cr
\+& $\bne\bne = \bo + \bfo + \bfi$    & $\bne\bz = \bt + \btr + \bsi$ \cr
\+& & \cr
\+& $\bz\bz = \bo + \bfo + \bfi$             & \cr
}\eqno{(A.2)}$$
One recognizes that the field $\bt$ is a `simple current' reflecting
the supersymmetric structure of this model. We conclude that the symmetry
algebra of this model can be extended by this `simple current' which
can be viewed as the inverse procedure of the projection
onto the bosonic part of the fermionic algebra.
Doing so we sum up the fields lying in one orbit under
$\bt$ and arrive at the following natural definitions:
$$\vbox{\settabs\+\indent&$\nso := {1\over2}
\bigl(\bfi+\bsi\bigr)$xxx&\hskip 1cm \cr
\+& $\nso := {1\over2}\bigl(\bo + \bt\bigr)$ & $\ro := \bse$        \cr
\+& $\nst := {1\over2}\bigl(\btr+\bfo\bigr)$ & $\rt := \ba$         \cr
\+& $\nstr:= {1\over2}\bigl(\bfi+\bsi\bigr)$
& $\rtr:= {1\over2}\bigl(\bne+\bz\bigr)$ \cr
}\eqno{(A.3)}$$
With this definition we recover exactly the fusion algebra $(A.1)$ of the
$N=1$ super Virasoro minimal model $c(2,12)$ from the fusion algebra $(A.2)$
of the $\w(2,4)$ rational model. This verifies
the consistency of the formulae presented in section $2$.
\bn
As a second example we consider the fusion \alg of the $N=1$ super
Virasoro minimal model $c(3,7)=-{11\over{14}}$. This model has three
HWRs per sector: $NS:h\in\{0,{2\over7},-{1\over{14}}\}$,
\hskip 0.5cm
$R:h\in\{{11\over{16}},-{3\over{112}},{13\over{122}}\}$.
There is no fixed point in the superconformal grid. With the
formulae of sections $2$ and $4$ we obtain the
following fusion algebra:
\def\nso{\cfs{NS}{0}}\def\nst{\cfs{NS}{{2\over7}}}
\def\nstr{\cfs{NS}{-{1\over{14}}}}\def\ro{\cfs{R}{{11\over{16}}}}
\def\rt{\cfs{R}{-{3\over{112}}}}\def\rtr{\cfs{R}{{13\over{122}}}}
$$\vbox{\settabs 2 \columns
\+ $\nst\nst = \nso +\nstr$            & $\nst\nstr = \nst + \nstr$    \cr
\+ $\nstr\nstr = \nso + \nst + \nstr$  &                               \cr
\+ & \cr
\+ $\ro\ro = 2\nso$                    & $\ro\rt = 2\nst$              \cr
\+ $\ro\rtr = 2\nstr$                  & $\rt\rt = 2\nso + 2\nstr$     \cr
\+ $\rt\rtr = 2\nst + 2\nstr$          & $\rtr\rtr=2\nso+2\nst+2\nstr$ \cr
}\eqno{(A.4)}$$
We already pointed out that this fusion algebra has a $\Zed_2$-structure.
As mentioned above the bosonic projection of the $N=1$ super Virasoro
\alg yields a $\w(2,4)$ for $c=-{11\over{14}}$. This value of the
central charge is also contained in the minimal series of the Virasoro
algebra and can be parametrized by $c=c_{Vir}(7,12)=-{11\over{14}}$.
The calculations in $\q{\wirrep}$ showed that $\w(2,4)$ has a
rational model at this value of $c$ and that the $\w$-characters
diagonalize the modular invariant partition function $(E_6,A_6)$ given by:
$$Z=\sum\limits^3_{s=1} \bigl(\mid\chi_{1,s}+\chi_{7,s}\mid^2+
\mid\chi_{4,s}+\chi_{8,s}\mid^2+\mid\chi_{5,s}+\chi_{11,s}\mid^2\bigr).$$
Hence the $\w(2,4)$ characters read:
$$\vbox{\settabs\+
$\chi^{\w,1}_s = \chi_{1,s}+\chi_{7,s}$\hskip 0.5cm &
$\chi^{\w,2}_s = \chi_{4,s}+\chi_{8,s}$\hskip 0.5cm &
$\chi^{\w,2}_s = \chi_{5,s}+\chi_{11,s}$\hskip 0.5cm & \cr
\+ $\chi^{\w,1}_s = \chi_{1,s}+\chi_{7,s}$ &
    $\chi^{\w,2}_s = \chi_{4,s}+\chi_{8,s}$ &
    $\chi^{\w,3}_s = \chi_{5,s}+\chi_{11,s}$ & $s=1,2,3\,.$ \cr
}\eqno{(A.5)}$$
Thus, this $\w(2,4)-$ minimal model 
has the following $9$ HWRs:\vskip 0.01pt\noindent
$h\in\{0,\dh,{2\over7},{11\over{14}},-{1\over{14}},{3\over7},{11\over{16}},
-{3\over{112}},{13\over{112}}\}$. Using the $S$-matrix for Virasoro minimal
models and the ordinary Verlinde formula one arrives $-$
after performing the change of basis $(A.5)$ $-$ at the following
fusion \alg for the $\w(2,4)$-minimal model $c=-{11\over{14}}$:
\mn
\def\bo{\cfs{}{0}}
\def\bt{\cfs{}{{3\over2}}}\def\btr{\cfs{}{2\over7}}
\def\bfo{\cfs{}{{11\over{14}}}}\def\bfi{\cfs{}{-{1\over{14}}}}
\def\bsi{\cfs{}{{3\over7}}}\def\bse{\cfs{}{{11\over{16}}}}
\def\ba{\cfs{}{-{3\over{112}}}}\def\bne{\cfs{}{{13\over{112}}}}
$$\vbox{\settabs\+xx&$\bfi\bfi=\bo+\btr+\bfo+\bfi$&\hskip 0.5cm\cr
\+& $\bt\bt = \bo$   & $\bt\btr = \bfo$ \cr
\+& $\bt\bfo = \btr$ & $\bt\bfi = \bsi$ \cr
\+& $\bt\bsi = \bfi$ & $\bt\bse = \bse$ \cr
\+& $\bt\ba = \ba$   & $\bt\bne = \bne$ \cr
\+& & \cr
\+& $\btr\btr = \bo + \bsi$   & $\btr\bfo =\bt + \bfi$  \cr
\+& $\btr\bfi = \bfo +\bfi$   & $\btr\bsi =\btr + \bsi$ \cr
\+& $\btr\bse = \ba$          & $\btr\ba  =\bse + \bne$ \cr
\+& $\btr\bne = \ba + \bne$   &  \cr
\+& & \cr
\+& $\bfo\bfo = \bo + \bsi$   & $\bfo\bfi =\btr + \bsi$ \cr
\+& $\bfo\bsi = \bfo + \bfi$  & $\bfo\bse = \ba$        \cr
\+& $\bfo\ba = \bse + \bne$   & $\bfo\bne = \ba + \bne$ \cr
\+& & \cr
\+& $\bfi\bfi= \bo + \btr + \bsi$ & $\bfi\bsi= \bt + \bfo + \bfi$ \cr
\+& $\bfi\bse = \bne$             & $\bfi\ba = \ba + \bne$        \cr
\+& $\bfi\bne = \bse +\ba + \bne$ &  \cr
\+& $\bsi\bsi=\bo + \btr + \bsi$ & $\bsi\bse = \bne$              \cr
\+& $\bsi\ba = \ba + \bne$       & $\bsi\bne = \bse + \ba + \bne$ \cr
\+ & & \cr
\+& $\bse\bse = \bo + \bt$    & $\bse\ba = \btr + \bfo$  \cr
\+& $\bse\bne = \bfi + \bsi$  & \cr
\+& & \cr
\+& $\ba\ba =\bo + \bt + \bfi + \bsi$  & \cr
\+& $\ba\bne = \btr+\bfo+\bfi+\bsi$    & \cr
\+& & \cr
\+& $\bne\bne = \bo + \bt + \btr + \bfo + \bfi + \bsi$  &  \cr
}\eqno{(A.6)}$$
Once again we recognize that the field $\bt$ is a `simple current'
which reflects the additional supersymmetry of this model.
Using the above conclusions and summing up the fields
belonging to the same orbit under the action of
$\bt$, we make the following definition:
$$\vbox{
\settabs\+\indent&$\nso:={1\over2}\bigl(\bfi+\bsi\bigr)$xxx&\hskip 1cm\cr
\+& $\nso := {1\over2}\bigl(\bo + \bt\bigr)$ & $\ro := \bse$  \cr
\+& $\nst := {1\over2}\bigl(\btr+\bfo\bigr)$ & $\rt := \ba$   \cr
\+& $\nstr:= {1\over2}\bigl(\bfi+\bsi\bigr)$ & $\rtr:= \bne$  \cr
}\eqno{(A.7)}$$
Using this definition we recover the fusion algebra $(A.4)$
from the fusion algebra $(A.6)$ of the $\w(2,4)$-minimal model.
\bn
\vfill\eject
\section{References}
\tolerance=10000
\mn
\bibitem{\bpz} A.A.\ Belavin, A.M.\ Polyakov, A.B.\ Zamolodchikov,
{\it Infinite Conformal Symmetry in Two-Dimensional
Quantum Field Theory,}
Nucl. Phys. {\bf B}241 (1984) p.\ 333
\bibitem{\witten} E.\ Witten, {\it Quantum Field Theory and the
Jones Polynomial,} Commun. Math. Phys. 121 (1989) p.\ 351
\bibitem{\cardy} J.L.\ Cardy, {\it Operator Content of Two-Dimensional
Conformally Invariant Theories,}
Nucl. Phys. {\bf B}271 (1986) p.\ 186
\bibitem{\ver} E.\ Verlinde, {\it Fusion Rules and Modular
Transformations in 2D Conformal Field Theory,}
Nucl. Phys. {\bf B}300 (1988) p.\ 360
\bibitem{\moresei} G.\ Moore, N.\ Seiberg,
{\it Polynomial Equations for Rational Conformal Field Theories,}
Phys. Lett. {\bf B}212 (1988) p.\ 451
\bibitem{\zamo} A.B.\ Zamolodchikov,
{\it Infinite Additional Symmetries in Two-Dimensional
Conformal Quantum Field}
{\it Theory}, Theor.\ Math.\ Phys.\ 65 (1986) p.\ 1205
\bibitem{\bai} F.A.\ Bais, P.\ Bouwknegt, M.\ Surridge, K.\ Schoutens,
{\it Coset Construction for Extended Virasoro Algebras,}
Nucl. Phys. {\bf B}304 (1988) p.\ 371
\bibitem{\blg} A.\ Bilal, J.L.\ Gervais,
{\it Systematic Construction of Conformal Theories with Higher-Spin
Virasoro Symmetries,}
Nucl.\ Phys.\ {\bf B}318 (1989) p.\ 579
\bibitem{\bal} J.\ Balog, L.\ Feh\'er, P.\ Forg\'acs,
L.\ O'Raifeartaigh, A.\ Wipf,
{\it Kac-Moody Realization of $\w$-Algebras,}
Phys.\ Lett.\ {\bf B}244 (1990) p.\ 435
\bibitem{\bou} P.\ Bouwknegt, {\it Extended Conformal Algebras},
Phys.\ Lett.\ {\bf B}207 (1988) p.\ 295
\bibitem{\blm} R.\ Blumenhagen, M.\ Flohr, A.\ Kliem, W.\ Nahm,
A.\ Recknagel, R.\ Varnhagen,
{\it $\w$-Algebras with Two and Three Generators},
Nucl.\ Phys.\ {\bf B}361 (1991) p.\ 255
\bibitem{\kauwat} H.G.\ Kausch, G.M.T.\ Watts,
{\it A Study of $\w$-Algebras Using Jacobi Identities,}
Nucl.\ Phys.\ {\bf B354} (1991) p.\ 740
\bibitem{\wattswb} G.M.T.\ Watts,
{\it WB Algebra Representation Theory,}
Nucl.\ Phys.\ {\bf B339} (1990) p.\ 177
\bibitem{\wattsrr} G.M.T.\ Watts,
{\it Null vectors of the superconformal algebra: the Ramond sector,}
preprint DAMTP-93-14, hep-th/9306034
\bibitem{\rva} R.\ Varnhagen, {\it Characters and Representations of
 New Fermionic $\w$-Algebras,}
Phys.\ Lett.\ {\bf B}275 (1992) p.\ 87
\bibitem{\wirrep} W.\ Eholzer, M.\ Flohr, A.\ Honecker,
R.\ H{\"u}bel, W.\ Nahm, R.\ Varnhagen,
{\it Representations of $\w$-Algebras with Two Generators
and New Rational Models}
Nucl.\ Phys.\ {\bf B383} (1992) p.\ 249
\bibitem{\kwf} E.\ Frenkel, V.\ Kac, M.\ Wakimoto,
{\it Characters and Fusion Rules for $\w$-Algebras via
Quantized Drinfeld-Sokolov Reduction,}
Commun.\ Math.\ Phys.\ 147 (1992) p.\ 295
\bibitem{\jos} J.M.\ Figueroa-O'Farrill, S.\ Schrans,
{\it The Conformal Bootstrap and Super-$\w$-Algebras},
Int.\ J.\ Mod.\ Phys.\ {\bf A}7 (1992) p.\ 591
\bibitem{\blmwir} R.\ Blumenhagen, W.\ Eholzer,
A.\ Honecker, R.\ H{\"u}bel,
{\it New $N=1$ Extended Superconformal Algebras
with Two and Three Generators},
Int.\ J.\ Mod.\ Phys.\ {\bf A}31 (1992) p.\ 7841
\bibitem{\eh1} W.\ Eholzer, A.\ Honecker, R.\ H{\"u}bel,
{\it Representations of N = 1 Extended Superconformal Algebras,}
Mod.\ Phys.\ Lett.\ {\bf A}8 (1993) p.\ 725
\bibitem{\sevrin} K.\ Schoutens, A.\ Sevrin,
{\it Minimal Super-$\w_N$ Algebras in Coset Conformal Field Theories,}
Nucl.\ Phys.\ {\bf B327} (1989) p.\ 673
\bibitem{\howcom} W.\ Eholzer, A.\ Honecker, R.\ H\"ubel,
{\it How Complete is the Classification of $\w$-Symmetries ?,}
Phys.\ Lett.\ {\bf B308} (1993) p.\ 42
\bibitem{\cas} M.\ Caselle, G.\ Ponzano, F.\ Ravanini,
{\it Towards a Classification of Fusion}
{\it Rule Algebras in Rational Conformal Field Theories,}
Int.\ J.\ Mod.\ Phys.\ {\bf B}6 (1992) p.\ 2075
\bibitem{\ehf} W.\ Eholzer, {\it Fusion Algebras Induced
by Representations of the Modular Group}, pre\-print BONN-HE-92-30,
to be published in Int.\ J.\ Mod.\ Phys.\ {\bf A}
\bibitem{\bouwschou} P.\ Bouwknegt, K.\ Schoutens,
{\it $\w$-Symmetry in Conformal Field Theory,}
Phys.\ Rep.\ {\bf 223} (1993) p.\ 183
\bibitem{\laszlo} L.\ Feh\'er, L.\ O'Raifeartaigh,
P.\ Ruelle and I.\ Tsutsui,
{\it On the Completeness of the Set of Classical $\w$-Algebras
Obtained from DS Reductions,}
preprint BONN--HE--93--14, DIAS--STP--93--02 (1993), hep-th/9304125
\bibitem{\dhr} S.\ Doplicher, R.\ Haag, J.E.\ Roberts
{\it Local Observables and Particle Statistics I, II,}
Commun.\ Math.\ Phys.\ 23 (1971) p.\ 199,
Commun.\ Math.\ Phys.\ 35 (1974) p.\ 49
\bibitem{\reck} A.\ Recknagel, {\it Fusion Rules from Algebraic K-Theory,}
Int.\ J.\ Mod.\ Phys.\ {\bf A}8 (1993) p.\ 1345
\bibitem{\www} W.\ Nahm, {\it A Proof of Modular Invariance,}
proceedings of the conference on `Topological Methods in
Quantum Field Theories',
ICTP, Trieste, Italy (1990) p.\ 75
\bibitem{\wowoB} W.\ Eholzer, N.\ Skoruppa, {\it Exceptional
$\w$-Algebra Characters
and Theta-Series of Quaternion Algebras}, in preparation
\bibitem{\mus} G.\ Mussardo, G.\ Sotkov, M.\ Stanishkov,
{\it Ramond Sector Structure of the Supersymmetric Minimal Models},
Phys.\ Lett.\ {\bf B}195 (1987) p.\ 397,
{\it Fine Structure of the Supersymmetric Operator Product
Expansion Algebras},
Nucl.\ Phys.\ {\bf B}305 (1988) p.\ 69
\bibitem{\schell} A.N.\ Schellekens, S.\ Yankielowicz,
{\it Simple Currents, Modular Invariants and Fixed Points,}
Int.\ J.\ Mod.\ Phys.\ {\bf A}5 (1990) p.\ 2903
\bibitem{\cap} A.\ Cappelli, C.\ Itzykson, J.B.\ Zuber,
{\it The A-D-E Classification of Minimal and
$A_1^{(1)}$ Conformal Invariant Theories,}
Commun.\ Math.\ Phys.\ 113 (1987) p.\ 1
\bibitem{\capsup} A.\ Cappelli, {\it Modular Invariant Partition Functions
of Superconformal Theories,}
Phys.\ Lett.\ {\bf B}185 (1987) p.\ 349
\bibitem{\Witten} E.\ Witten, {\it Constraints on Supersymmetry Breaking,}
Nucl.\ Phys.\ {\bf B}202 (1982) p.\ 253
\bibitem{\Kastor} D.\ Kastor, {\it Modular Invariance
in Superconformal Models,}
Nucl.\ Phys.\ {\bf B}280 (1987) p.\ 304
\bibitem{\pim} M.\ Flohr, {\it $\w$-Algebras, New Rational Models and
Completeness of the $c=1$}
{\it Classification}, preprint BONN-HE-92-08 (1992),
to be published in Commun.\ Math.\ Phys.
\bibitem{\roc} A.\ Rocha-Caridi, {\it Vacuum Vector
Representations of the Virasoro Algebra,}
in Vertex Operators in Mathematics and Physics (1984)
S.\ Mandelstam and I.M.\ Singer, eds., p.\ 451
\bibitem{\roes} J.\ Kellendonk, M.\ R\"osgen, R.\ Varnhagen,
{\it  Path Spaces and $\w$-Fusion in Minimal Models,}
preprint BONN-HE-93-04, to be published in Mod.\ Phys.\ Lett.\ {\bf A}
\bibitem{\goddard} P.\ Goddard, A.\ Kent, D.\ Olive,
{\it Unitary Representations of the Virasoro and Super-Virasoro Algebras,}
Commun.\ Math.\ Phys.\ 103 (1986) p.\ 105
\bibitem{\matsuo} Y.\ Matsuo, S.\ Yahikozawa, {\it Superconformal
Field Theories with
Modular Invariance on a Torus},
Phys.\ Lett.\ {\bf B}178 (1986) p.\ 211
\bibitem{\Ravanini} F.\ Ravanini, S.\ Yang, {\it Modular Invariance in
$N=2$ Superconformal Field Theories,}
Phys.\ Lett.\ {\bf B}195 (1987) p.\ 202
\bibitem{\mussSS} G.\ Mussardo, G.\ Sotkov, M.\ Stanishkov,
{\it $N=2$ Superconformal Minimal Models,}
Int.\ J.\ Mod.\ Phys.\ {\bf A}5 (1989) p.\ 1135
\bibitem{\ralfdip} R.\ H\"ubel, {\it Darstellungstheorie von
$\w$- und Super-$\w$-Algebren,}
Diplomarbeit, BONN-IR-92-11
\bibitem{\bowknegt} P.\ Bouwknegt, {\it Extended Conformal
Algebras from Kac-Moody Algebras,}
Proceedings of the meeting `Infinite dimensional Lie algebras and Groups'
CIRM, Luminy, Marseille (1988) p.\ 527
\bibitem{\horstphd} H.G.\ Kausch, {\it Chiral Algebras in
Conformal Field Theory,}
Ph.D.\ thesis, Cambridge University, September 1991
\bibitem{\andicorr} A.\ Honecker, {\it A Note on the
Algebraic Evaluation of Correlators
in Local Chiral Conformal Field Theory},
preprint BONN-HE-92-25 (1992), hep-th/9209029
\bibitem{\pimpriv} M.\ Flohr, private communication
\vfill
\end